\documentclass[useAMS, usenatbib, a4paper]{mnras}
\usepackage[spanish,es-minimal,english]{babel}
\usepackage[utf8]{inputenc}
\usepackage{graphicx}
\usepackage{xcolor}
\usepackage{fixltx2e}
\usepackage{hyperref}
\usepackage{savesym}
\savesymbol{tablenum}
\usepackage{siunitx}
\restoresymbol{SIX}{tablenum}
\usepackage{newtxtext}
\usepackage[varg,varvw,smallerops]{newtxmath}
\usepackage{xfrac} 
\usepackage{booktabs}
\usepackage{array}   
\newcolumntype{L}{>{$}l<{$}} 
\newcolumntype{R}{>{$}r<{$}} 
\newcolumntype{C}{>{$}c<{$}}
\hypersetup{colorlinks=True, linkcolor=blue!50!black, citecolor=black,
  urlcolor=blue!50!black}

\usepackage{etoolbox}
\robustify\bfseries
\robustify\itshape
\usepackage{enumerate}

\DeclareSIUnit\msun{\text{M\ensuremath{_\odot}}}
\DeclareSIUnit\lsun{\text{L\ensuremath{_\odot}}}

\newcounter{ionstage}
\renewcommand{\ion}[2]{\setcounter{ionstage}{#2}%
  \ensuremath{\mathrm{#1\,\scriptstyle\Roman{ionstage}}}}
\newcommand\hii{\ion{H}{2}}

\def\th#1#2{\ensuremath{\theta^{#1}\,\text{Ori~#2}}}
\newcommand\wn{\ensuremath{\tilde{\nu}}}

\newcommand*\chem[1]{\ensuremath{\mathrm{#1}}}
\newcommand\Config[1]{\ensuremath{\mathrm{#1}}}
\newcommand\Term[3]{\ensuremath{\mathrm{#1\ ^{#2}#3}}}
\newcommand\Level[4]{\ensuremath{\mathrm{#1\ ^{#2}#3_{#4}}}}

\newcommand\ha{\ensuremath{\text{H}\alpha}}
\newcommand\hb{\ensuremath{\text{H}\beta}}

\newcommand\lyb{\ensuremath{\text{Ly}\beta}}

\newcommand\Raman{\ensuremath{_{\text{Raman}}}}
\newcommand\scat{\ensuremath{_{\text{scat}}}}

\newcommand\wing{\ensuremath{_{\text{wing}}}}
\newcommand\lamcont{\ensuremath{_{\lambda, \text{cont}}}}
\newcommand\observed{\ensuremath{^{\text{obs}}}}

\title[Raman mapping of PDRs]
{Raman mapping of photodissociation regions}

\author[Henney]{
  William J. Henney\thanks{w.henney@irya.unam.mx}\\
  \foreignlanguage{spanish}{Instituto de Radioastronomía y
    Astrofísica, Universidad Nacional Autónoma de México, Apartado
    Postal 3-72, 58090 Morelia, Michaoacán, Mexico}}

\date{Accepted XXX. Received YYY; in original form ZZZ}

\pubyear{2019}

\begin{document}
\label{firstpage}
\pagerange{\pageref{firstpage}--\pageref{lastpage}}
\maketitle


\begin{abstract}
  Broad Raman-scattered wings of hydrogen lines can be used to
  map neutral gas illuminated by high-mass stars in star forming regions.
  Raman scattering transforms far-ultraviolet starlight
  from the wings of the \lyb{} line (\SI{1022}{\angstrom} to \SI{1029}{\angstrom})
  to red visual light in the wings of the \ha{} line
  (\SI{6400}{\angstrom} to \SI{6700}{\angstrom}).
  Analysis of spatially resolved spectra of the Orion Bar and other regions
  in the Orion Nebula shows that this process occurs in
  the neutral photo-dissociation region between the ionization front and dissociation front.
  The inner Raman wings are optically thick and allow the neutral hydrogen density
  to be determined, implying  \(n(\chem{H^0}) \approx \SI{e5}{cm^{-3}}\) for the Orion Bar.
  Far-ultraviolet resonance lines of neutral oxygen imprint their absorption
  onto the stellar continuum as it passes through the ionization front,
  producing characteristic absorption lines
  at \SI{6633}{\angstrom} and \SI{6664}{\angstrom} with widths of order \SI{2}{\angstrom}.
  This is a unique signature of Raman scattering, which allows it
  to be easily distinguished from other processes that might produce broad \ha{} wings,
  such as electron scattering or high-velocity outflows.
\end{abstract}
\section{Introduction}
\label{sec:introduction}

\begin{figure}
  \centering
  \includegraphics[width=\linewidth]{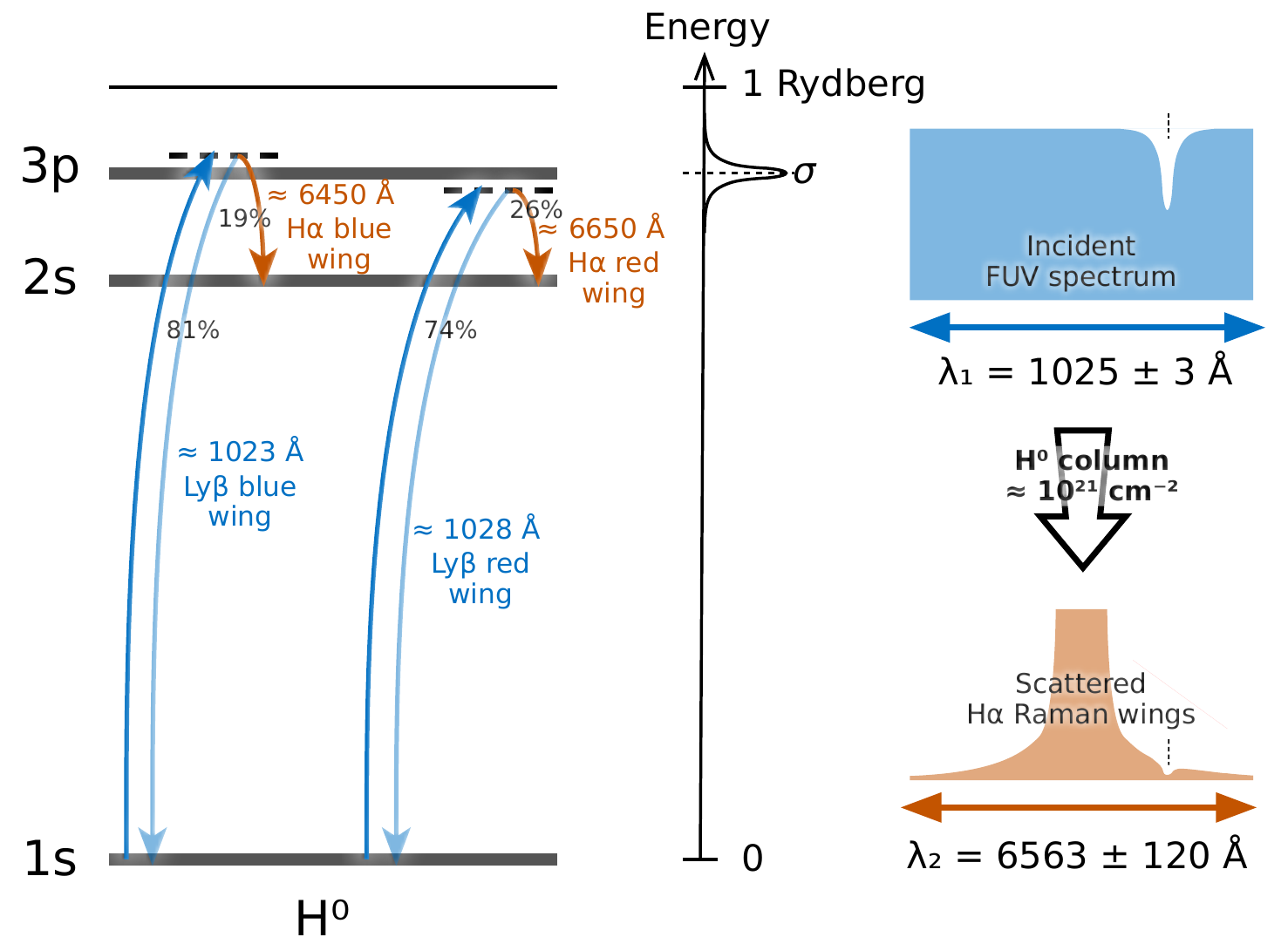}
  \caption{Schematic illustration of Raman scattering of photons from
    the \lyb{} wings to the \ha{} wings.  The relevant energy levels
    of neutral hydrogen are shown at left.  Far ultraviolet photons
    that are shifted by about \(\Delta\lambda_1 = \pm 1\) to \SI{3}{\angstrom}
    from the \lyb{} rest wavelength can excite transitions from the
    ground \Config{1s} level to a virtual state adjacent to
    \Config{3p}.  Most such excitations decay back to \Config{1s}
    (Rayleigh scattering), but in about one-fifth of cases the decay
    is to \Config{2s} instead (Raman scattering).  The scattering
    cross section falls approximately as \(\Delta\lambda_1^{-2}\), which gives
    broad Lorentzian wings to the \ha{} line, as shown at right. A
    bandwidth of \(\Delta\lambda_1 = \pm \SI{3}{\angstrom}\) around \lyb{} is
    transformed to a bandwidth
    \(\Delta\lambda_2 \approx \pm \SI{120}{\angstrom}\) around \ha{}.  A narrow
    absorption line in the incident FUV spectrum (vertical thin dashed
    line) becomes a much broader notch in the scattered wings. }
  \label{fig:raman-cartoon}
\end{figure}

Raman scattering is the inelastic analog of Rayleigh scattering by
atoms or molecules.  Both processes begin with a radiation-induced
transition of an electron to a virtual bound state (non-eigenstate).
In Rayleigh scattering, the electron returns to its original state,
resulting in the radiation being re-emitted with its original
frequency (elastic scattering).  In Raman scattering, on the other
hand, the electron undergoes a transition to a different excited
state, resulting in radiation being re-emitted at a much lower
frequency.  See Figure~\ref{fig:raman-cartoon} for an illustration of
the process. Recently, \citet{Dopita:2016a} identified exceedingly
broad wings to the \ha{} \SI{6563}{\angstrom} line in the Orion Nebula
and a number of \hii{} regions in the Magellanic Clouds, which they
ascribe to Raman scattering of ultraviolet radiation in the vicinity
of the \lyb{} \SI{1025}{\angstrom} transition.  Raman scattering in
astrophysical sources was first identified in symbiotic stars
\citep{Schmid:1989a}, where FUV \ion{O}{6} emission lines at
\num{1032} and \SI{1038}{\angstrom} produce broad emission features at
\num{6827} and \SI{7088}{\angstrom}.  This illustrates a curious
feature of Raman scattering \citep{Nussbaumer:1989a}: the relative
width \(\Delta\lambda/\lambda\) of spectral features is amplified by a factor
\(\lambda(\ha)/\lambda(\lyb) \approx 6.4\) when passing from the FUV to the optical
domain.

\citet{Dopita:2016a} propose that the Raman wings form at the
transition zone near the ionization fronts in \hii{} regions.
However, the total neutral hydrogen column through the ionization
front can be no more than about
\(10 / \sigma_0 \approx \SI{2e18}{cm^{-2}}\), where
\(\sigma_0 \approx \SI{6.3e-18}{cm^2}\) is the ground-state hydrogen
photoionization cross section at threshold \citep{Osterbrock:2006a}.
The Raman scattering cross section at wavelengths responsible for the
observed wings is much lower than this:
\(\sigma\Raman \sim \SI{e-21}{cm^2}\) \citep{Chang:2015a}, meaning that the
Raman scattering optical depth through the ionization front is only of
order \(0.001\).  A vastly larger column density of neutral hydrogen
(\(\approx \SI{e21}{cm^{-2}}\)) is available in the photodissociation region
(PDR) outside the ionization front, so it is more likely that Raman
scattering will occur there instead, so long as there is sufficient
far ultraviolet radiative flux.

This paper is organized as follows.
Section~\ref{sec:muse-spectr-mapp} presents archival VLT-MUSE
integral field spectroscopy of the Orion Nebula.
Spectra of the broad H\(\alpha\) wings are extracted for key areas of the nebula
(section~\ref{sec:raman-scattered-ha}).
Two components of the \ion{O}{1} UV resonance multiplet
\(\Term{2s 2p^4}{3}{P} \to \Term{2s 2p^3 3d}{3}{D^o}\) are detected as absorption
features at \num{6633} and \SI{6664}{\angstrom} against the \ha{}
Raman wings (section~\ref{sec:raman-scattered-fuv}).
Three wavelength bands are defined in each of the red and blue wings
that avoid contaminating emission and absorption features (section~\ref{sec:defin-observ-bands}).
The emission bands are spatially mapped and compared with
other tracers of ionized and neutral zones in the nebula
(section~\ref{sec:spat-mapp-raman}).
Particular attention is paid to the edge-on photodissociation region at the Orion Bar
(section~\ref{sec:emiss-prof-across}).
Equivalent widths of the Raman-scattered absorption lines are compared with
the distribution of other absorption features in the nebula
(section~\ref{sec:equiv-widths-absorpt}).
Section~\ref{sec:keck-observations} presents archival
Keck-HIRES slit spectroscopy, which shows the profile of the
\SI{6664}{\angstrom} absorption line with an effective velocity
resolution of \SI{1}{km.s^{-1}}.
Section~\ref{sec:discussion} discusses
the implications of these results for the structure of
the PDRs in Orion.
An appendix 
recapitulates the basic theory of Raman scattering, concentrating on
the wavelength transformation from the FUV domain around \lyb{} to
optical domain around \ha{}.  In addition, polynomial fits are
provided to the wavelength dependence of the total (Rayleigh plus
Raman) scattering cross section and the Raman \ha{} branching
ratio.

\section{Spectral mapping of Raman wings}
\label{sec:muse-spectr-mapp}
\begin{figure}
  \centering
  \includegraphics[width=\linewidth]{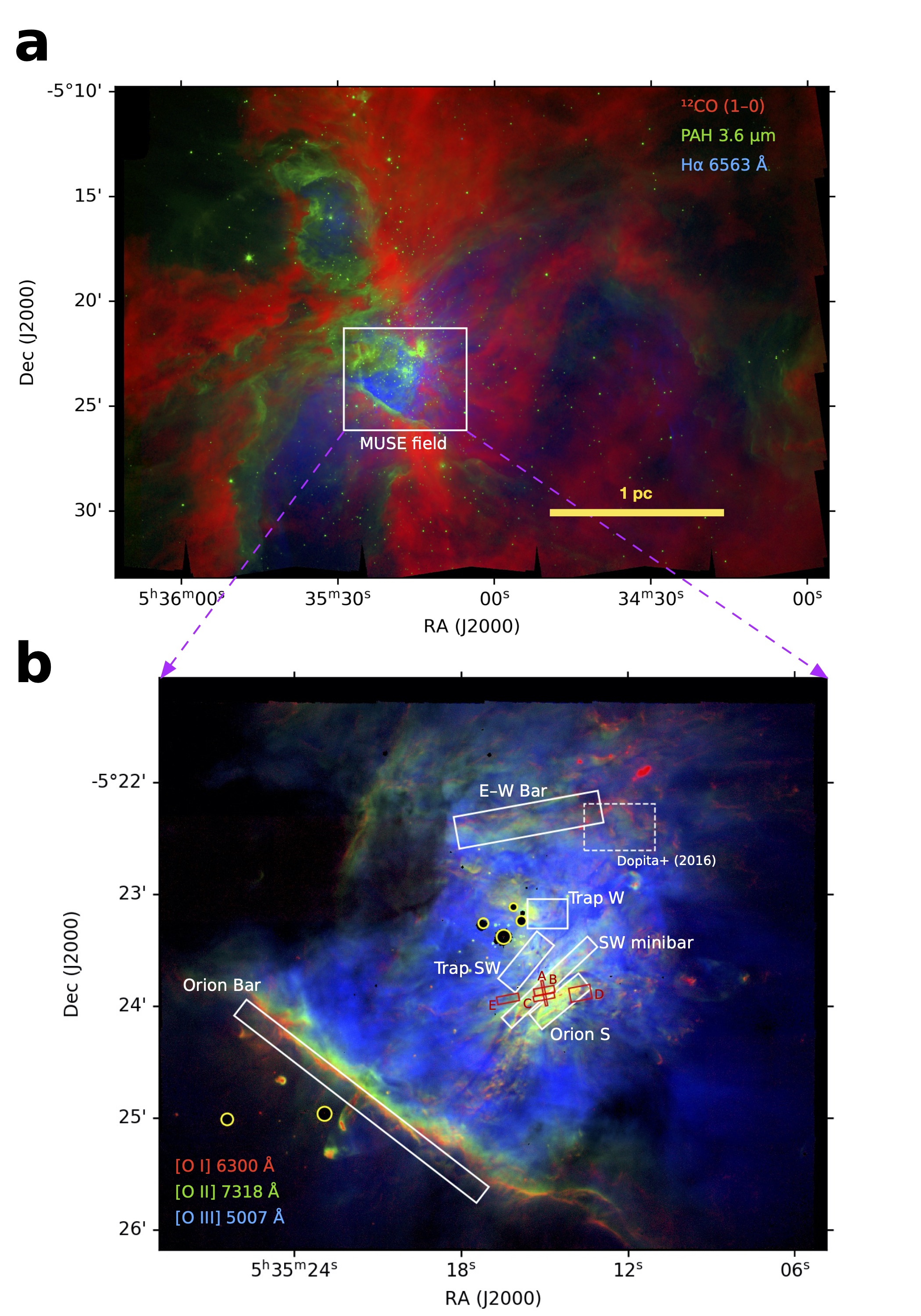}
  \caption{(a)~Panoramic view of the Orion Nebula region on parsec
    scales.  Red channel of background image shows \chem{^{18}CO}
    emission from dense molecular gas \citetext{Carma-NRO Orion
      Survey, \citealp{Kong:2018a}}. Green channel shows near-infrared
    continuum from PAH grains in neutral gas \citetext{Spitzer Orion
      Survey, \citealp{Megeath:2012a}}.  Blue channel shows optical
    H\(\alpha\) emission from ionized gas \citetext{WFI camera on ESO
      \SI{2.2}{m} La Silla, \citealp{Da-Rio:2009a}}. The white box
    shows the field of view observed with MUSE.  (b)~Locations within
    the nebula of the extraction regions for the MUSE spectra shown in
    Figure~\ref{fig:raman-spectra-1d} (white boxes) and the Keck HIRES
    spectra shown in Figure~\ref{fig:raman-keck} (red boxes).  The
    region of the spectrum obtained in \citet{Dopita:2016a} is shown
    by the dashed box. Background image shows oxygen emission lines
    extracted from the MUSE data cube: [\ion{O}{1}] \(\lambda6300\) (red
    channel), which traces ionization fronts and shocks in neutral
    gas; [\ion{O}{2}] \(\lambda7318\) (green channel), which traces the
    outer layers of the photoionized nebula; and [\ion{O}{3}]
    \(\lambda5007\) (blue channel), which traces the highly ionized interior
    of the nebula. }
  \label{fig:raman-fov-regions}
\end{figure}
The principal observational dataset used in this paper is the imaging
spectroscopy mosaic of the inner Orion Nebula \citep{Weilbacher:2015a,
  Mc-Leod:2016a} obtained with the MUSE spectrograph
\citep{Bacon:2010a, Bacon:2014a} on the VLT.  The entire datacube
covers the wavelength range \num{4595} to \SI{9366}{\angstrom} but we
concentrate mainly on the range \num{6300} to \SI{6800}{\angstrom},
where the spectral resolving power is \(R \approx 2500\), corresponding to
an instrumental linewidth (FWHM) of
\(\delta\lambda \approx \SI{2.4}{\angstrom}\), which is sampled at
\SI{1.25}{\angstrom.pix^{-1}} and then resampled to
\SI{0.85}{\angstrom.pix^{-1}} for the final calibrated cube
\citetext{see \S~2 of \citealp{Weilbacher:2015a}}.  The observed field
is shown by the white rectangle in
Figure~\ref{fig:raman-fov-regions}(a) and includes the entire inner
Huygens region of the nebula, which accounts for roughly half of the
total radio continuum flux from the \hii{} region
\citep{Subrahmanyan:2001a}.  The full datacube is a mosaic that
combines observations from 30 separate pointings, with a pixel size of
\(0.2 \times 0.2''\).

\subsection{Raman scattered \boldmath\ha{} wings}
\label{sec:raman-scattered-ha}

\begin{figure*}
  \includegraphics[width=\linewidth]{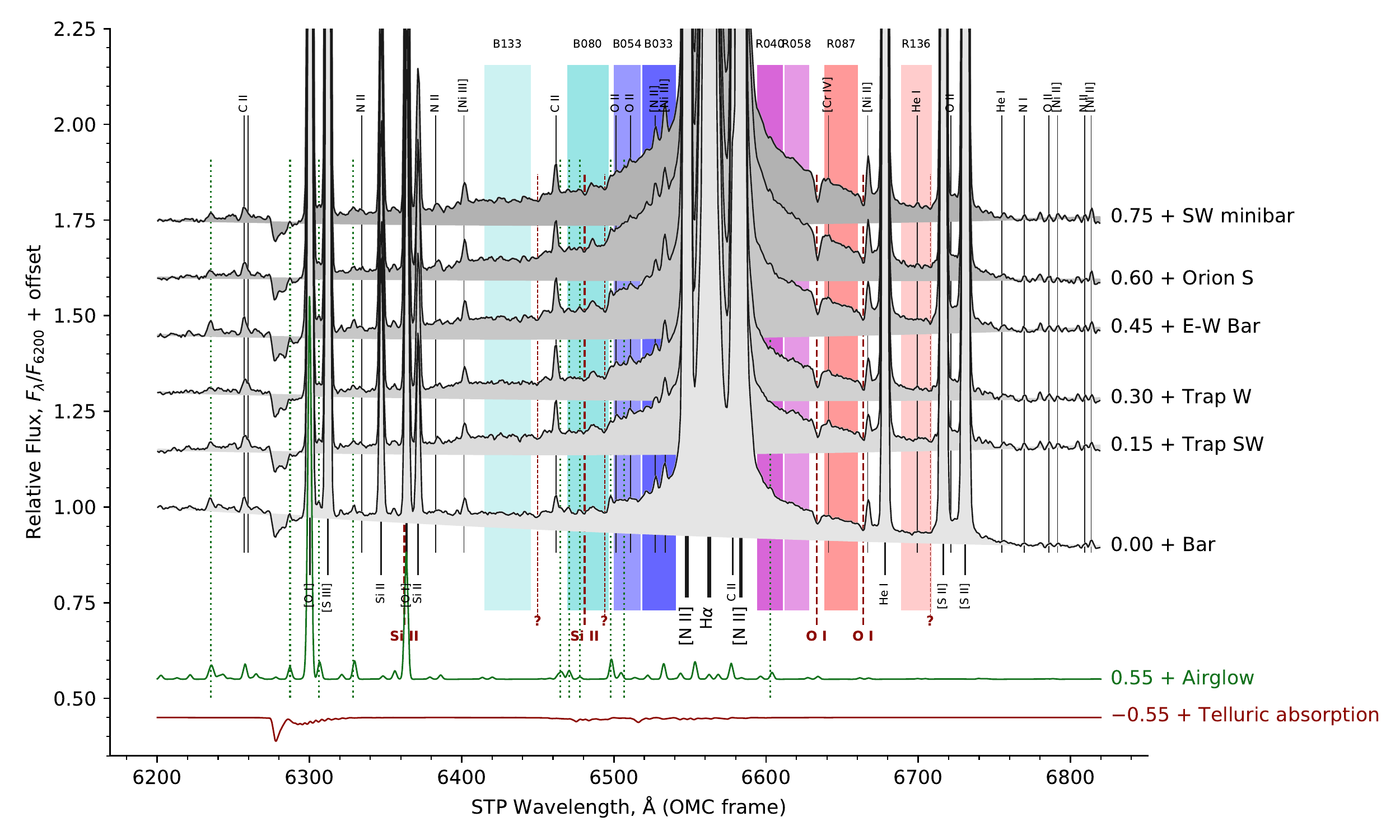}
  \caption{MUSE spectra of the Orion Nebula centered on the
    H\(\alpha\) line, showing the broad Raman-scattered wings.  Spectra are
    shown for the regions outlined by white boxes in
    Figure~\ref{fig:raman-fov-regions} and were extracted from the
    datacubes of \citet{Weilbacher:2015a}. All spectra are normalized
    to the continuum at \SI{6200}{\angstrom} and are offset vertically
    as labelled at right. Emission line identifications are shown by
    solid vertical lines, labelled above for weak lines and below for
    stronger lines.  Wavelengths of Raman-scattered \ion{O}{1} and
    \ion{Si}{2} absorption lines from
    Table~\ref{tab:raman-wavelengths} are shown by red dashed vertical
    lines, labelled in bold font below, together with unidentified
    candidate absorption features marked with ``?''.  Dotted lines
    show night sky lines identified in the Dark Bay region of the
    nebula.  At the bottom of the figure are shown the theoretical
    upper-atmosphere airglow spectrum (green line) and telluric
    molecular absorption spectrum (red line), as calculated from the
    Cerro Paranal Advanced Sky Model \citep{Noll:2012a,
      Moehler:2014a}.  }
  \label{fig:raman-spectra-1d}
\end{figure*}

Extended wings to the \ha{} line are detected over the entire map, but
they are particularly prominent in the six regions marked as white
boxes in Figure~\ref{fig:raman-fov-regions}(b).  These are all
bar-like features \citep{ODell:2000a, Garcia-Diaz:2007a}, which
correspond to filamentary ionization fronts.  Extracted MUSE spectra
around the \ha{} line for each of these regions are shown in
Figure~\ref{fig:raman-spectra-1d}.  All regions show broad wings to
the \ha{} line extending from \num{6300} to \SI{6700}{\angstrom},
which are consistent with the higher spectral resolution
(\(R = 7000\)) results of \citet{Dopita:2016a} for the region
indicated by a dashed box in Figure~\ref{fig:raman-fov-regions}(b).
The continuum is interpolated under the wings by fitting a
second-order polynomial to clean regions of the spectra in the ranges
\SIrange{6070}{6225}{\angstrom} and \SIrange{6760}{6820}{\angstrom},
as shown by the lower edge of the gray shading for each spectrum.  The
continuum is a combination of Paschen recombination emission from
hydrogen, which increases to the red, and dust-scattered starlight and
two-photon hydrogen emission, which both increase to the blue.  The
combined \(F_\lambda\) is approximately flat for all regions except the
Orion Bar, where dust scattering of the nearby star \th2A{} dominates,
resulting in a blueward rise.  All of these continuum processes are
expected to be smooth on a \SI{100}{\angstrom} scale,\footnote{The
  only exception is the dust-scattered \ha{} line from the stellar
  photospheres and winds, but this is confined within
  \(\pm \SI{30}{\angstrom}\) of the rest wavelength (see Fig.~2 of
  \citealp{Simon-Diaz:2006b}) and so does not affect the Raman wings,
  which are broader than that.}  but nonetheless the systematic
uncertainties in the continuum interpolation is an important factor
that limits the precision of the Raman wing measurements.

\subsection{Raman scattered FUV absorption lines}
\label{sec:raman-scattered-fuv}
A prominent feature of all the spectra in
Figure~\ref{fig:raman-spectra-1d} is a pair of absorption notches in
the red wing at \SI{6633}{\angstrom} and \SI{6664}{\angstrom}, which
closely correspond to the wavelengths expected for the Raman-scattered
transformation of the \ion{O}{1} resonance lines at
\SI{1027}{\angstrom} and \SI{1028}{\angstrom} listed in
Table~\ref{tab:raman-wavelengths}.  After correcting for instrumental
broadening, the FWHM of the \SI{6633}{\angstrom} feature is
\(\approx \SI{2.5}{\angstrom}\), which would correspond to a velocity width
of \SI{115}{km.s^{-1}} if it were an optical absorption line.
However, taking into account the wavelength transformation during
Raman scattering (eq.~[\ref{eq:wav-transform}]), the true velocity
width of the FUV line is \SI{18}{km.s^{-1}}.  The absorption depth of
the features is roughly 50\% with respect to the Raman-scattered wing
at adjacent wavelengths, but is much less than this with respect to the total
continuum emission.  The same spectral features are also visible in
the \citet{Dopita:2016a} spectra but are not commented on by those
authors. The \SI{6664}{\angstrom} feature is partially blended with
the [\ion{Ni}{2}] \SI{6666.8}{\angstrom} nebular emission line in the
MUSE spectra, but these are better resolved in the higher resolution
\citeauthor{Dopita:2016a} spectra.  Archival Keck HIRES observations
of this feature at an even higher spectral resolution are presented
below in \S~\ref{sec:keck-observations}.  The identification of these
two FUV absorption lines in the optical spectrum is incontestable
proof of a Raman scattering origin for the broad \ha{} wings.

There is also evidence for a weak absorption feature in the blue wing
at \SI{6481}{\angstrom}, corresponding to the \ion{Si}{2} absorption
line at \SI{1024}{\angstrom}.  However, the blue wing is much less
clean than the red wing, partly due to telluric absorption and airglow
emission (see below), which makes the identification uncertain.  The
Raman wings show some other weak features that remain unidentified,
the strongest of which is an apparent broad
(\(\approx \SI{5}{\angstrom}\)) absorption feature at \SI{6494}{\angstrom},
while additional narrower features are seen at \SI{6450}{\angstrom}
and \SI{6708}{\angstrom} (all marked with ``?'' in
Figure~\ref{fig:raman-spectra-1d}).  It is not clear if these are
truly absorption features or whether they are simply gaps between very
weak blended emission lines.

\subsection{Definition of observational Raman bands in the red and blue wings}
\label{sec:defin-observ-bands}

\begin{table*}
  \caption{Wavelength bands used for Raman wing extraction}
  \label{tab:wav-bands}
  ~\\[-\baselineskip]
  \centering
  \begin{tabular}{l l R C C C C l}\toprule
    & Band & \langle\Delta\lambda_2\rangle, \si{\angstrom}   & \lambda_{\text{min}}, \si{\angstrom} & \lambda_{\text{max}}, \si{\angstrom} & f_{\mathrm{H\alpha}} & \langle\sigma_\lambda\rangle, \SI{e-21}{cm^2} & Contamination \\
    & (1) & (2) & (3) & (4) & (5) & (6) & (7) \\
    \midrule
    Blue wing & B133 & -132.8& 6414.85& 6445.45 & 0.180 & 0.144 & \\
    & B080 & -79.5 & 6469.25& 6496.45 & 0.196 & 0.378 & Sky 6471, 6478, \chem{H_2 O}\\
    & B054 & -53.6 & 6499.85& 6517.70 & 0.205 & 0.802 & \ion{O}{2}? 6502, 6510, Sky 6507, \chem{H_2 O}\\
    & B033 & -32.8 & 6518.55& 6540.65 & 0.212 & 2.069 & [\ion{N}{2}] 6527.24, [\ion{Ni}{3}] 6533.76, \chem{H_2 O}\\
    \addlinespace[2pt]
    Red wing & R040 & 40.3 &  6594.20& 6611.20 & 0.238 & 1.451 & Sky 6603\\
    & R058 & 57.7 &  6612.05& 6628.20 & 0.244 & 0.695 & \\
    & R087 & 87.1 &  6638.40& 6660.50 & 0.255 & 0.299 & [\ion{Cr}{4}]? 6641\\
    & R136 & 135.7 & 6688.55& 6708.95 & 0.274 & 0.119 & \ion{He}{1} 6699\\
    \bottomrule
    \multicolumn{8}{p{15cm}}{
    \textsc{Columns:} (1)~Name of band.  (2)~Mean wavelength displacement from \ha{} rest wavelength. (3, 4)~Upper and lower wavelength limits for the band. (5)~Mean value of the \ha{} branching ratio from virtual \Config{3p} levels (see equation~[\ref{eq:fha-fit}]).  (6)~Mean value of \lyb{} wing cross-section that can feed this band via Raman scattering (see equation~[\ref{eq:total-cross-section-fit}]).  (7)~Nebular and telluric lines that may contaminate the band (see text for details).
    }
  \end{tabular}
\end{table*}

In order to study the spatial distribution of the Raman-scattered
wings, it is convenient to define a series of broad bands on the blue
and red sides, which are listed in Table~\ref{tab:wav-bands} and shown
as blue and red shaded vertical stripes in
Figure~\ref{fig:raman-spectra-1d}. Four bands are defined in each
wing, spanning a range in \(|\Delta\lambda_2|\) from about \SI{30}{\angstrom} to
\SI{150}{\angstrom}.  The lower limit of this range is determined by
overlap with the strong nebular [\ion{N}{2}] emission lines, while the
upper limit is due the [\ion{S}{2}] lines on the red side, combined
with the Raman wings becoming too weak to measure.  The mean
\chem{H^0} \lyb{} cross section corresponding to each band
(eq.~[\ref{eq:total-cross-section-fit}]) is also given in the table,
varying between about \SI{1e-22}{cm^2} and \SI{2e-21}{cm^2}.

The bands are chosen so as to avoid the strongest contaminating lines
wherever possible, but some small contamination is unavoidable, as
listed in the rightmost column of the table.  The contamination comes
from two sources: weak nebular emission lines (indicated by solid
vertical lines in Figure~\ref{fig:raman-spectra-1d}) and additionally
from the line absorption and emission of the Earth's atmosphere.  Two
complementary methods were used to investigate this latter effect.
First, the datacube was inspected to identify lines with roughly
uniform brightness across the entire map.  In particular, any line
that is as strong in the Dark Bay region as it is in Orion~S is
unlikely to come from the nebula itself.  Such lines are indicated by
vertical dotted lines in Figure~\ref{fig:raman-spectra-1d} and some
are listed in Table~\ref{tab:wav-bands}.  Second, ESO's SkyCalc
tool\footnote{\url{http://www.eso.org/sci/software/pipelines/skytools/skycalc}}
was used to predict theoretical emission and absorption spectra for
the atmosphere above the VLT at the time and airmass of the
observations, convolved with the MUSE instrumental profile.  The
emission spectrum is dominated by upper-atmosphere airglow lines of
\chem{OH} \citep{Osterbrock:1996a, Noll:2012a, Noll:2014a}, shown in
green in the figure, many of which can be seen to coincide with the
empirically determined sky lines.  The absorption spectrum is
dominated by telluric lines of \chem{O_2} and \chem{H_2 O}
\citep{Moehler:2014a, Smette:2015a}.  The strongest predicted
absorption near \ha{} is the \chem{O_2} \(\gamma\) band at
\SI{6280}{\angstrom}, which is clearly seen in all the spectra, with
an absorption depth of order \num{0.1}, but lies well away from the
Raman wings.  The blue Raman wing is affected by weaker \chem{H_2 O}
bands at \SIrange{6460}{6600}{\angstrom} with absorption depth
\(< 0.01\), as compared to the relative brightness of the Raman wings
at those wavelengths, which is \numrange{0.05}{0.1}.

In summary, the nebular and telluric contamination introduces
uncertainties of order 10\% in the fluxes of the B033, B054, and B080
bands, with other bands being affected to a much lesser degree.

\subsection{Spatial distribution of Raman band emission}
\label{sec:spat-mapp-raman}
\begin{figure*}
  \includegraphics[width=\linewidth]{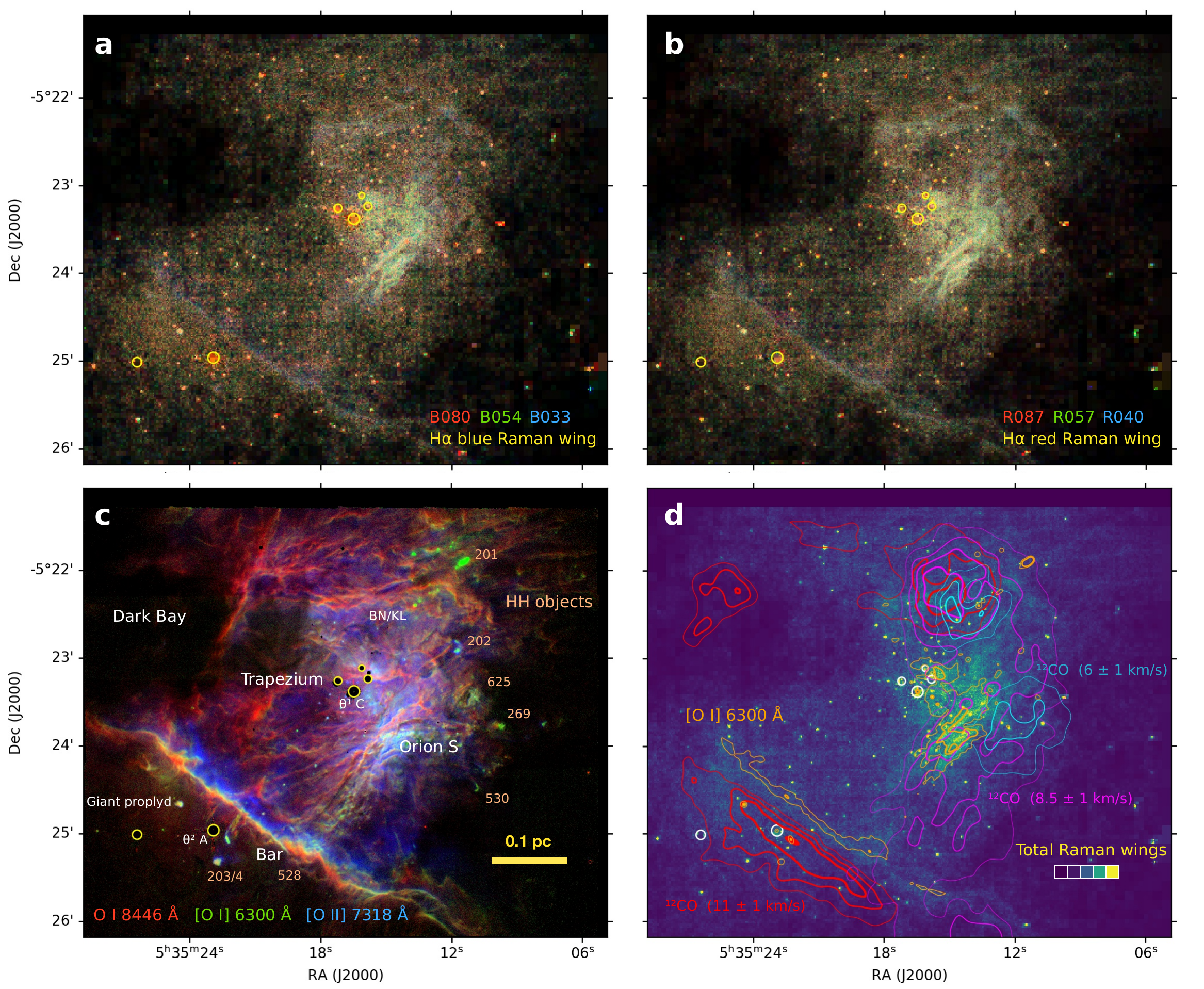}
  \caption{Spatial distribution of Raman-scattered wings of
    H\(\alpha\). (a)~Three-color RGB image of bands in the blue wing: B080
    (red channel), B054 (green), B033 (blue).  All bands are
    continuum-subtracted and are multibinned at constant
    signal-to-noise (see text).  Maximum brightness in the (red,
    green, blue) channels is
    \((2.27, 3.89, 7.78) \times
    \SI{1e-9}{erg.s^{-1}.cm^{-2}.sr^{-1}.\angstrom^{-1}}\). The
    brightest stars are masked out and marked with yellow circles.
    Lower-mass stars are not masked and most appear as yellow/red
    dots. (b)~Same as (a) but for bands in the red wing: R087 (red),
    R057 (green), R040 (blue). (c)~Three-color RGB image of
    collisional and fluorescent oxygen emission lines that bracket the
    ionization front: \ion{O}{1} \(\lambda8446\) (red), [\ion{O}{1}]
    \(\lambda6300\) (green), [\ion{O}{2}] \(\lambda7318\) (blue).  Various
    features of the nebula are marked, including Herbig-Haro objects
    driven by jets from young stars.  (d)~Comparison of the total
    Raman wing intensity (blue-to-yellow false color image) with the
    [\ion{O}{1}] emission that traces the ionization front (orange
    contours) and \chem{^{12}CO} emission that traces molecular gas
    (red, purple, and blue contours for different LSR velocities as
    marked).}
  \label{fig:raman-maps}
\end{figure*}

Figure~\ref{fig:raman-maps} shows maps of the six innermost Raman
bands as two RGB images: the blue wing in panel~a and the red wing in
panel~b.  In each case, the channel sequence R, G, B is an inward
progression towards line center, from the farther wings to the nearer
wings.  The maps are adaptively smoothed following the binary grid
algorithm outlined in \citet{Garcia-Diaz:2018a}, which reduces the
noise in the fainter regions at the cost of a reduced spatial
resolution.  The two maps are strikingly similar, indicating that the
contamination of the blue wing (see previous section) is indeed a
minor effect when integrated over an entire band.

To aid in the interpretation of the Raman wing maps,
Figure~\ref{fig:raman-maps}c shows a combined view of
mid-to-low-ionization oxygen emission lines, derived from the MUSE
data cube.  In blue is shown the [\ion{O}{2}]
\(7318.39, \SI{7319.99}{\angstrom}\) doublet, which traces the outer
10\% of the fully ionized emission.  In green is shown the
[\ion{O}{1}] \SI{6300.30}{\angstrom} line, which principally traces
partially ionized gas at the ionization front, but also shocks in
neutral gas. In red is shown the fluorescent \ion{O}{1}
\SI{8446.36}{\angstrom} line, which traces fully neutral gas that is
very close behind the ionization front.  This image clearly
illustrates that the neutral/molecular gas is organized in filaments,
with bright ionization fronts on the side facing the high-mass
Trapezium stars. This is most clearly seen in the Orion Bar, the
linear emission feature to the south-east of the map, but analogous
filaments are seen in all directions from the Trapezium (see also
Figure~\ref{fig:raman-fov-regions}b).  A particularly complex region,
with several partially overlapping filaments, lies between the
Trapezium and the Orion~S star formation region.  This region shows
the brightest Raman wing emission, suggesting that it contains the
neutral gas with the highest incident FUV flux, presumably because it
lies physically closest to the illuminating high-mass stars.

The spatial relation of the Raman-scattered wings to other emission
lines is illustrated in Figure~\ref{fig:raman-maps}d, which shows the
summed wing intensity over all 8 bands as a false color scale. This is
compared with contours that show the ionization front, as traced by
the collisionally excited [\ion{O}{1}] line, and molecular gas, as
traced by the optically thick \chem{^{12}CO} (1--0) line
\citep{Kong:2018a}.  In the Bar region, the Raman emission is clearly
seen to be sandwiched between the ionization front and the molecular
gas, conclusive evidence that it arises in the neutral zone of the
PDR.\@ In the Orion~S region and around the Trapezium, there is not
such a clear stratification between ionized, neutral, and molecular
emission.  This is due to two factors: first, the densities are
higher, which shortens all length scales, and second, the geometry is
not so edge-on, leading to a greater degree of superposition along the
line of sight, as witnessed by the overlap of the [\ion{O}{1}] and
\chem{^{12}CO} contours.  Nonetheless, even here there is evidence
that the Raman emission tends to lie farther from \th1C{} than the
ionization front, particularly in the ``SW minibar'' region.

\subsection{Emission profiles across the Orion Bar}
\label{sec:emiss-prof-across}

\begin{figure*}
  \includegraphics[width=\linewidth]{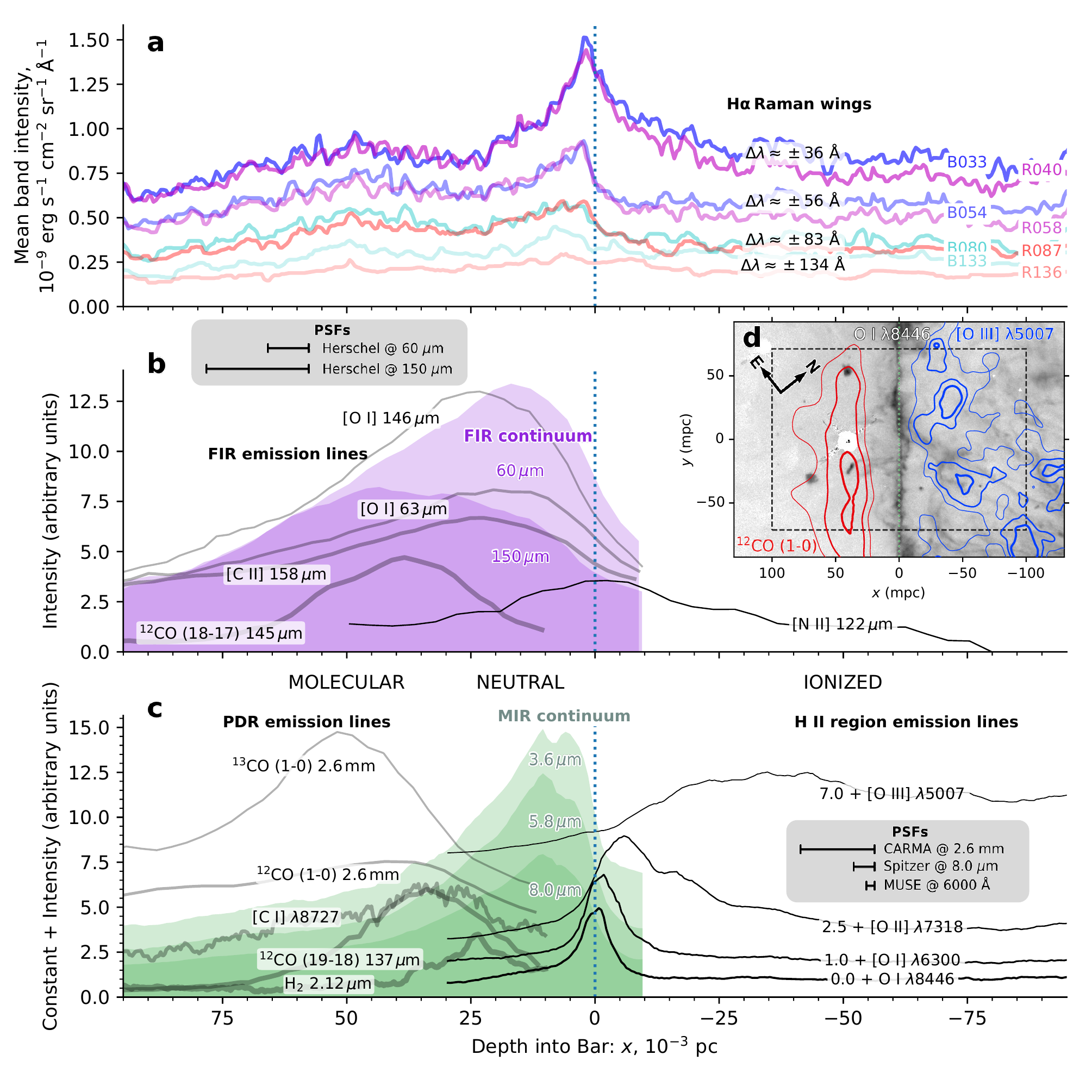}
  \caption{(a)~Spatial cut of continuum-subtracted Raman-scattered
    H\(\alpha\) wing intensity across the Orion Bar. Matched band pairs
    from the red and blue wings are shown with the same color scheme
    as in Fig.~\ref{fig:raman-spectra-1d}. (b)~Comparison with
    far-infrared observations of line and continuum (purple shading)
    emission from Herschel PACS \citep{Bernard-Salas:2012a}.  The
    width (FWHM) of the Herschel PSF is shown by horizontal bars for
    two representative wavelengths. (c)~Comparison with a variety of
    other tracers of molecular, neutral, and ionized material.
    Optical lines are from VLT MUSE \citep{Weilbacher:2015a}, infrared
    continuum bands (green shading) are from Spitzer IRAC
    \citep{Megeath:2012a}, CO lines are from Carma-NRO
    \citep{Kong:2018a} and Herschel PACS \citep{Parikka:2018a},
    \chem{H_2} near-infrared line is from VLT HAWK-I
    \citep{Kissler-Patig:2008a}. (d)~Image of the bar in \ion{O}{1},
    rotated so the \(x\) axis is along the direction of the spatial
    cuts (area indicated by dashed rectangle).  Contours of molecular
    CO emission (red) and ionized [\ion{O}{3}] emission (blue) are
    also shown.  In all four panels, ultraviolet photons are incident
    from the right hand side and the vertical dotted line indicates
    the ionization front, defined as the peak of the fluorescent
    \ion{O}{1} emission. The profiles are the median values over a
    slit width of \(75''\), centered on the ICRS equatorial
    coordinates \(\alpha = \ang{83.8419}\), \(\delta = \ang{-5.4113}\) and
    oriented at a position angle of \ang{141.5}.  Distances from the
    ionization front are given in \si{mpc} where
    \(1'' \approx \SI{1.9}{mpc}\) for an assumed distance of \SI{388 \pm
      5}{pc} \citep{Kounkel:2017a}.}
  \label{fig:raman-bar-profile}
\end{figure*}

\begin{table*}
  \newlength\tablegap\setlength\tablegap{6pt}
  \caption{Peaks in emission profiles across the Orion Bar}
  \label{tab:bar-profile-peaks}
  ~\\[-\baselineskip]
  \centering
  \begin{tabular}{r r R R R R c }\toprule
    Emission type and units
    & Line or Band & x_0, \si{mpc} & \delta x, \si{mpc} & I(\text{Peak}) & I(\text{BG}) & Ref.\\
    (1) & (2) & (3) & (4) & (5) & (6) & (7)\\
    \midrule
    \hii{} region emission lines
    & [\ion{O}{3}] \(\lambda 5007\) & -32.1 \pm 2.6& 45 & 13.059 & 11.642 \pm 5.605 & 1 \\
    & [\ion{O}{2}] \(\lambda 7318\) & -9.4 \pm 3.6& 25 & 0.882 & 0.232 \pm 0.026 & 1 \\    
    Peak and BG: \smash{\(\int I_\lambda \, d\lambda\), \SI{e-6}{erg.s^{-1}.cm^{-2}.sr^{-1}}}
    & [\ion{O}{1}] \(\lambda 6300\) & -1.7 \pm 0.1& 9 & 0.152 & 0.045 \pm 0.001 & 1 \\     
    & \ion{O}{1}   \(\lambda 8446\) & 0.0 \pm 0.8& 8 & 0.226 & 0.071 \pm 0.000 & 2 \\      
    \addlinespace[\tablegap]
    \ha{} Raman bands
    & B033 & 0.8 \pm 1.5 & 20 & 0.455 & 0.858 \pm 0.007 & 2 \\
    & R040 & 1.8 \pm 0.1 & 18 & 0.480 & 0.815 \pm 0.023 & \\
    Peak and BG: \smash{\(I_\lambda\), \SI{e-9}{erg.s^{-1}.cm^{-2}.sr^{-1}.\angstrom^{-1}}} 
    & B054 & 5.0 \pm 2.7 & 9 & 0.233 & 0.641 \pm 0.009  & \\
    & R058 & 5.5 \pm 2.8 & 11 & 0.285 & 0.591 \pm 0.017 & \\
    & B080 & 7.1 \pm 1.3 & 11 & 0.136 & 0.437 \pm 0.013 & \\
    & R087 & 7.1 \pm 0.7 & 15 & 0.178 & 0.393 \pm 0.017 & \\
    & B133 & 6.7 \pm 1.1 & 10 & 0.128 & 0.302 \pm 0.004 & \\
    & R136 & 3.1 \pm 2.7 & 27 & 0.072 & 0.198 \pm 0.007 & \\
    \addlinespace[\tablegap]
    Spitzer IRAC bands
    & \SI{3.6}{\micron} & 11.6 \pm 1.1 & 21 & 0.200 & 0.182 \pm 0.018 & 3 \\
    Peak and BG: \(I_\nu\), \si{GJy.sr^{-1}}                              
    & \SI{5.8}{\micron} & 12.8 \pm 2.4 & 23 & 1.702 & 0.966 \pm 0.073 & \\
    & \SI{8.0}{\micron} & 13.9 \pm 3.4 & 24 & 4.866 & 2.622 \pm 0.119 & \\
    \addlinespace[\tablegap]
    Herschel PACS continuum
    & \SI{60}{\micron} &  23.9 \pm 7.0 & 52 & 66.361 & 25.175 \pm 0.261 & 4\\
    Peak and BG: \(I_\nu\), \si{Jy.pix^{-1}}
    & \SI{150}{\micron} & 39.2 \pm 4.2& 63 & 7.263 & 2.592 \pm 0.065 & \\
    \addlinespace[\tablegap]
    PDR emission lines
    & [\ion{N}{2}] \SI{122}{\micron} & -4.0 \pm 1.7 & 55 & 1.907 & 0.581 \pm 0.181 & 4\\
    & [\ion{O}{1}] \SI{63}{\micron}  & 21.6 \pm 1.2 & 55 & 1.303 & 0.823 \pm 0.242 & 4\\
    Peak and BG: \smash{\(\int I_\lambda \, d\lambda\)}, median normalized
    & \chem{H_2} \SI{2.12}{\micron} & 23.6 \pm 0.1 & 19 & 0.169 & 0.997 \pm 0.002 & 5 \\
    & [\ion{C}{2}] \SI{158}{\micron} & 24.3 \pm 1.9 & 47 & 0.975 & 0.920 \pm 0.145 & 4\\
    & [\ion{O}{1}] \SI{146}{\micron} & 26.0 \pm 2.8 & 46 & 2.299 & 0.945 \pm 0.210 & 4\\
    & \chem{^{12}CO} (19--18) & 33.8 \pm 0.1 & 32 & 3.930 & 0.399 \pm 0.001 & 6 \\
    & [\ion{C}{1}] \(\lambda 8727\) & 37.7 \pm 5.9 & 30 & 0.495 & 0.854 \pm 0.122 & 2 \\
    & \chem{^{12}CO} (18--17) \SI{145}{\micron}  & 38.9 \pm 0.5 & 34 & 8.193 & 0.989 \pm 0.154 & 4 \\
    & \chem{^{12}CO} (1--0) & 41.1 \pm 1.2 & 44 & 0.472 & 0.746 \pm 0.173 & 7 \\
    & \chem{^{13}CO} (1--0) & 49.8 \pm 2.1 & 33 & 0.891 & 0.823 \pm 0.112 & 7 \\
    \bottomrule
    \multicolumn{7}{@{}p{16cm}@{}}{%
    \textsc{Note}---Results of fitting a single Gaussian plus linear
    background to the brightness profiles shown in
    Figure~\ref{fig:raman-bar-profile}.
    \textsc{Columns}: (1)~Type of tracer and surface brightness units for the values
    given in columns~5 and~6.
    (2)~Specific emission line or continuum band.
    (3)~Peak position of fitted Gaussian relative to ionization front.
    Negative values are on the ionized side of the front,
    positive values on the neutral/molecular side.
    The uncertainty is determined as the difference between the fitted Gaussian peak
    and the actual data peak.
    (4)~FWHM of fitted Gaussian. Note that this is not corrected for
    the instrumental PSF, which contributes at the 10--20\% level in
    the case of the Herschel and CARMA
    data (references~4 and~7) but is negligible in other cases.
    (5)~Peak intensity of fitted Gaussian, with brightness units given in column~1.
    (6)~Intensity of fitted background at position \(x_0 \pm \delta x\).
    (7)~Data provenance references:
    1.~VLT--MUSE \citep{Weilbacher:2015a};
    2.~VLT--MUSE (this paper);
    3.~Spitzer--IRAC \citep{Megeath:2012a};
    4.~Herschel--PACS \citep{Bernard-Salas:2012a};
    5.~VLT--HAWK-I \citep{Kissler-Patig:2008a};
    6.~Herschel--PACS \citep{Parikka:2018a};
    7.~Carma--NRO \citep{Kong:2018a}.
    }
  \end{tabular}
\end{table*}

Figure~\ref{fig:raman-bar-profile} shows spatial profiles across the
Orion Bar of the Raman-scattered bands (panel a), together with other
emission lines and bands (panels b and c). In keeping with the
dominant tradition in the literature \citetext{e.g., Fig.~9 of
  \citealp{van-der-Werf:1996a}, Fig.~2 of \citealp{Goicoechea:2017a}},
the molecular regions are shown on the left and the ionized regions on
the right.  The profiles are the median values across a slit of width
\(75''\), as illustrated in Figure~\ref{fig:raman-bar-profile}d.
Table~\ref{tab:bar-profile-peaks} shows the positions, widths, and
intensities of the peaks in each tracer, as determined by fitting a
single Gaussian plus a linear background.

\subsubsection{Lines from ionized gas and the ionization front at the Bar}
\label{sec:lines-from-ionized}

The ionization stratification at the edge of the \ion{H}{2} region is
clearly seen in the distribution of the [\ion{O}{3}], [\ion{O}{2}],
[\ion{O}{1}], and \ion{O}{1} lines
(Fig.~\ref{fig:raman-bar-profile}c).  The peak in [\ion{O}{1}]
emissivity (excited by collisions between electrons and neutral oxygen
atoms) is expected to occur at a hydrogen ionization fraction of 50\%
\citep{Henney:2005b}, which is displaced by
\(x_0 \approx \SI{-1.7}{mpc}\) from the peak of the \ion{O}{1} fluorescent
line.  This latter should correspond to an absorption optical depth of
approximately unity in the FUV pumping lines, such as those listed in
Table~\ref{tab:raman-wavelengths} \citetext{see \S~5 of
  \citealp{Walmsley:2000a}}, corresponding to a neutral hydrogen
column density of \(N(\chem{H^0}) \approx \SI{1e19}{cm^{-2}}\) and therefore
an average volume density of
\(n(\chem{H^0}) = N(\chem{H^0}) / |x_0| \approx \SI{2000}{cm^{-3}}\).  This
value is similar to the peak electron density on the fully ionized
side of the ionization front: \(n_e \approx \SI{3000}{cm^{-3}}\), measured
from the [\ion{S}{2}] \(6717/6731\) ratio \citep[e.g.][]{ODell:2017b},
which occurs at \(x \approx \SI{-7}{mpc}\), close to the peak in the
[\ion{O}{2}] emission. Therefore, the \emph{total} hydrogen density is
roughly constant over the transition between a predominantly ionized
state and predominantly neutral state.  Note that the apparent width
of this transition (for instance, the FWHM of the [\ion{O}{1}] peak
\(\approx \SI{9}{mpc}\)) is much larger than predicted by atomic physics,
which is probably due to spatial irregularities in the ionization
front, which can be seen in the inset
Figure~\ref{fig:raman-bar-profile}d.

\subsubsection{Spatial profiles of the Raman bands in the neutral Bar}
\label{sec:spat-prof-raman}
Unlike the optical emission lines, the Raman wing bands (upper panel
of Fig.~\ref{fig:raman-bar-profile}) all show peaks at positive values
of \(x\), corresponding to fully neutral gas in the PDR.\@ There is a
very close agreement between corresponding pairs of blue and red
bands: B033 with R040, B054 with R058, and B080 with R087.  However,
there is a systematic tendency for the blue band intensity to be
slightly higher than its red counterpart on the ionized side (negative
values of \(x\)), which is probably due to contamination by nebular
emission lines, as listed in column~7 of
Table~\ref{tab:wav-bands}. There is a clear tendency for the peak in
the brightness profile to progress towards greater depths into the
neutral gas as one moves to wavelengths farther from the \ha{} line
center, which is quantitatively confirmed by the Gaussian fits in
Table~\ref{tab:bar-profile-peaks}.
This result is used as a diagnostic of the PDR density in \S~\ref{sec:raman-scattering-as} below.

\subsubsection{Other PDR tracers of the neutral and molecular Bar}
\label{sec:other-pdr-tracers}

\newcommand\vibro[3]{\ensuremath{v = #1 \to #2\ \mathrm{#3}}}

The left hand side of panels b and c of
Figure~\ref{fig:raman-bar-profile} show a variety of other emission
lines and continuum bands, which trace different layers in the PDR.\@
The near-infrared \chem{H_2} \vibro{1}{0}{S(1)} line marks the
hydrogen dissociation front at \(x_0 \approx \SI{24}{mpc}\)
\citep{van-der-Werf:1996a, Luhman:1998a, Marconi:1998a},
which provides the outer boundary of the neutral hydrogen layer.
Beyond this are are found the high-\(J\) CO lines,
which trace the \chem{CO} dissociation front at \(x_0 \approx \SI{36}{mpc}\).
These coincide with near-infrared [\ion{C}{1}] emission,
which mainly traces the recombination of \chem{C^+} ions
in cool gas near the dissociation front \citep{Escalante:1991a}.
The mm-band CO lines trace the deeper, fully molecular regions,
with the optically thick \chem{^{12}CO\ (1-0)} line at \(x_0 \approx \SI{41}{mpc}\)
arising outside of the optically thin \chem{^{13}CO\ (1-0)} line
at  \(x_0 \approx \SI{50}{mpc}\)
\citep{Kong:2018a}. 

The broad infrared bands are dominated by dust emission.
For the case of the \num{3.6} to \SI{8.0}{\micro m}
\textit{Spitzer} bands \citep{Megeath:2012a},
mid-infrared spectroscopy \citep{Bregman:1989a, Cesarsky:2000a, Kassis:2006a}
shows that they are mainly due to discrete polycyclic aromatic hydrocarbon (PAH)
features at \num{3.3}, \num{6.2}, \num{7.7}, and \SI{8.6}{\micro m},
plus a continuum contribution from very small grains
(VSGs, with size \(a \sim \SI{0.007}{\micro m}\), \citealp{Desert:1990a})
and minor contributions from ionized gas lines such as [\ion{Ar}{2}] and [\ion{Ar}{3}].
The longer wavelength \num{60} and \SI{150}{\micro m} \textit{Herschel} bands
\citep{Bernard-Salas:2012a}
show continuum emission from larger grains (\(a \sim \SI{0.1}{\micro m}\)).
The peak of the dust emission moves deeper into the PDR with increasing wavelength,
reflecting the decreasing dust temperature as the stellar radiation field is attenuated
\citep{Arab:2012a}. 

A complementary view of the Bar is provided by far-infrared emission lines
of neutral and ionized metals
observed by \textit{Herschel} \citep{Bernard-Salas:2012a}.
The [\ion{N}{2}] \SI{122}{\micro m} line comes from the \chem{H^+} region
and shows a similar distribution to the optical [\ion{O}{2}] line.
The [\ion{C}{2}] \SI{158}{\micro m} and [\ion{O}{1}] \num{63} and \SI{146}{\micro m} lines,
on the other hand, come from the \chem{H^0} region of the PDR.


\begin{figure*}
  \includegraphics[width=\linewidth]{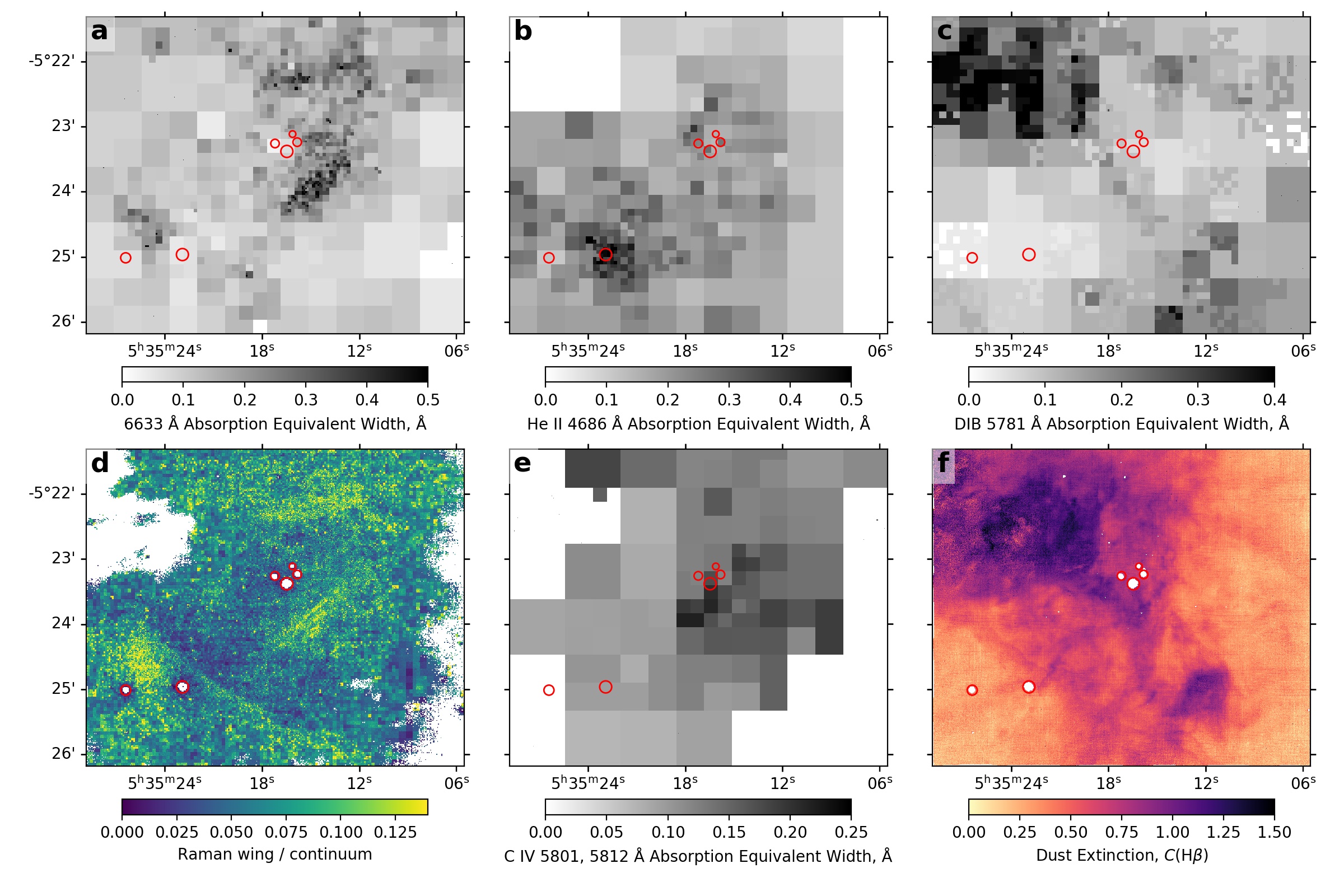}
  \caption{
    Maps of equivalent widths \(W_\lambda\observed\)
    of different absorption features (grayscale images),
    compared with maps of other quantities (false color images).
    Red circles show positions of bright stars,
    from West to East: \th1A, B, C, D, and \th2A, B.\@
    (a)~Equivalent width of \SI{6633}{\angstrom} absorption feature.
    (b)~Equivalent width of \ion{He}{2} \SI{4685.68}{\angstrom} absorption line.
    (c)~Equivalent width of a solid-state Diffuse Interstellar Band
    absorption feature at \SI{5781}{\angstrom}.
    (d)~Ratio \(R\wing\) of Raman wing intensity (R058 band)
    to the intensity of the underlying continuum at the same wavelength.
    (e)~Equivalent width of \ion{C}{4} \num{5801.35}, \SI{5811.97}{\angstrom}
    absorption line doublet (sum of the two lines).
    (f)~Foreground dust extinction to the ionized gas,
    \(C(\hb)\) (base-10 logarithmic extinction at \SI{4861}{\angstrom}),
    calculated from the Balmer decrement \(\ha{}/\hb{}\),
    assuming the reddening curve of \citet{Blagrave:2007a}.
  }
  \label{fig:raman-multi-absorption-features}
\end{figure*}

\subsection{Equivalent widths of absorption features}
\label{sec:equiv-widths-absorpt}

Figure~\ref{fig:raman-multi-absorption-features} shows grayscale maps
of the observed equivalent widths, \(W_\lambda\observed\),
of various absorption features that are found
superimposed on the continuous spectrum of the Orion Nebula.
These are extracted from the MUSE datacubes by integrating over
a short wavelength interval, \(\Delta\lambda = \pm \SI{8}{\angstrom}\),
around the central wavelength of the absorption feature:
\begin{equation}
  \label{eq:equivalent-width}
  W_\lambda\observed = \int_{\Delta\lambda}  \frac{I\lamcont - I_\lambda}{I\lamcont} \, d\lambda
  \quad \si{\angstrom},
\end{equation}
where \(I\lamcont\) is the continuum intensity,
which is interpolated from the absorption-free portions of the \(\Delta\lambda\) interval.

\subsubsection{Raman-scattered \ion{O}{1} \SI{1027}{\angstrom}}
\label{sec:cand-raman-scatt}

Figure~\ref{fig:raman-multi-absorption-features}a shows
\(W_\lambda\observed\) for the \SI{6633}{\angstrom} absorption feature,
which is a candidate for being a Raman-scattered FUV absorption line.
Note that in this case, the ``continuum'' intensity \(I\lamcont\)
includes a contribution from the broad Raman wing,
as well as the ``true'' nebular continuum and dust-scattered starlight.
Due to the weakness of both the continuum and the absorption line,
it was necessary to aggressively bin the data
in order to produce an acceptable map.
The results in Figure\ref{fig:raman-multi-absorption-features}d
are for a minimum S/N of 7 in \(W_\lambda\observed\), which gives
the optimum balance between noise and spatial resolution.

The 6633 equivalent width is non-zero across the entire map,
with a median value of  \(W_\lambda\observed (6633) \approx \SI{0.1}{\angstrom}\). 
It shows peaks of \(W_\lambda\observed (6633) = \SI{0.3 \pm 0.1}{\angstrom}\)
in exactly the same areas as the peaks in the Raman band emission
(see \S~\ref{sec:spat-mapp-raman}).
For ease of comparison,
Figure~\ref{fig:raman-multi-absorption-features}d shows a map of
the wing-to-continuum ratio, \(R\wing\),
for the R087 red Raman wing band, 
which is found by dividing 
the continuum-subtracted intensity in the MUSE data cube
by the intensity of the interpolated continuum:
\begin{equation}
  \label{eq:wing-ratio}
  R\wing = 
  \frac{
    \int_{\lambda_{\text{min}}}^{\lambda_{\text{max}}} \bigl(  I_\lambda - I'\lamcont \bigr) \, d\lambda 
  }{
    \int_{\lambda_{\text{min}}}^{\lambda_{\text{max}}}  I'\lamcont \, d\lambda 
  } , 
\end{equation}
where the band limits \(\lambda_{\text{min}}\), \(\lambda_{\text{max}}\)
are given in Table~\ref{tab:wav-bands}.
In this case, the continuum intensity \(I'\lamcont\)
does \emph{not} include the Raman wing emission 
since it is interpolated across
a much wider range than in equation~\eqref{eq:equivalent-width}
by fitting a quadratic function to clean sections of continuum around
\SI{6150 \pm 80}{\angstrom} and \SI{6790 \pm 30}{\angstrom}
(see \S~\ref{sec:raman-scattered-ha}).
Note also that \(R\wing\) is a dimensionless ratio,
whereas \(W_\lambda\observed\) has dimensions of wavelength.
The reason for choosing the R087 band for this figure is that it is
the band that is closest in wavelength to the \SI{6633}{\angstrom} absorption feature.
It is clear from comparing Figure~\ref{fig:raman-spectra-1d} panels~a and~d
that \(W_\lambda\observed (6633)\) and \(R\wing\) closely track one another,
as would be expected if the absorption feature is the Raman-scattered \ion{O}{1} \SI{1027}{\angstrom} line.

\begin{figure}
  \centering
  \includegraphics[width=\linewidth]{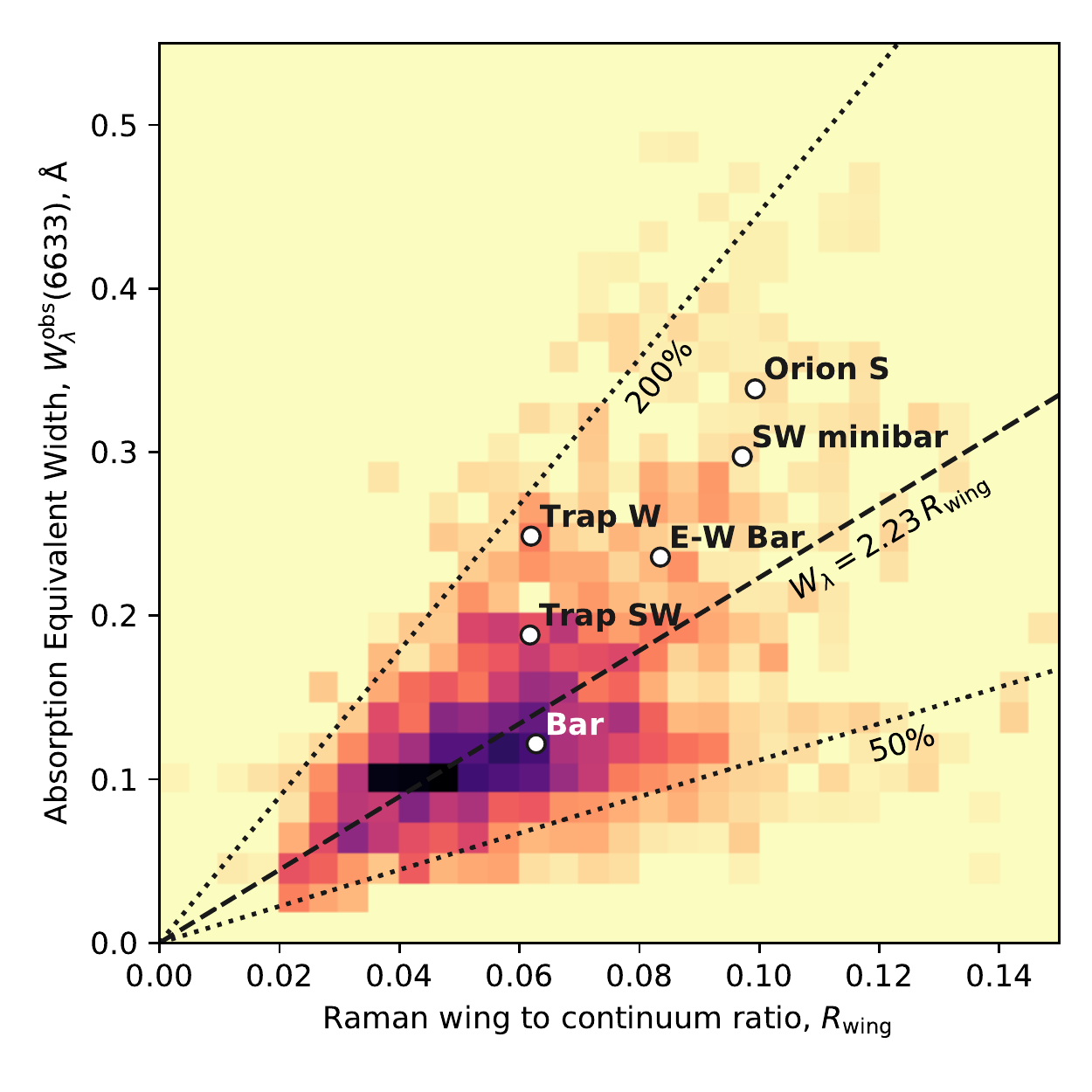}
  \caption{
    Correlation between \SI{6633}{\angstrom} equivalent width,
    \(W_\lambda\observed(6633)\), and Raman wing to continuum ratio, \(R\wing\).
    False-color image shows histogram of pixels across the face of the nebula,
    calculated from the MUSE spectral cube: 
    Fig.~\ref{fig:raman-multi-absorption-features} panel~a (\(y\) axis)
    and panel~d (\(x\) axis).
    The histogram is  weighted by the \SI{6630}{\angstrom}
    continuum intensity.
    The straight dashed line shows the mean relation,
    \(W_\lambda = 2.27\, R\wing\),
    while dotted lines show a \(2 \times\) steeper and \(2 \times\) shallower slope.
    Symbols show average values for each of the extraction regions
    shown by white boxes in Figure~\ref{fig:raman-fov-regions}b.
  }
  \label{fig:6633-Rwing}
\end{figure}

The relationship is examined more quantitatively in Figure~\ref{fig:6633-Rwing},
which shows that \(W_\lambda\observed (6633)\) and \(R\wing\) are well-correlated,
with a mean slope of \(W_\lambda = 2.27\, R\wing\).
Of the six extraction regions used in Figure~\ref{fig:raman-spectra-1d},
only the Orion Bar lies on this mean relation.
The other regions, which are closer to the Trapezium
(see Figure~\ref{fig:raman-fov-regions}b)
follow a slope that is roughly 50\% steeper.
Although this might represent a real physical difference,
it is also true that the lower-left area of the graph in Figure~\ref{fig:6633-Rwing}
is more susceptible to systematic errors.
For instance, in the Orion Bar region much of the continuum comes from
the blue reflection nebula around \th2A, which causes increased spectral curvature
(see Figure~\ref{fig:raman-spectra-1d}),
potentially leading to reduced accuracy in the continuum interpolation,
which might result in an overestimation of \(R\wing\).
Higher values of \(R\wing\) and \(W_\lambda\), such as those seen in Orion~S,
are much less affected by such uncertainties.

\subsubsection{Stellar absorption lines}
\label{sec:stell-absorpt-lines}

Figure~\ref{fig:raman-multi-absorption-features}b and~e show \(W_\lambda\observed\)
for the \ion{He}{2} \SI{4685.68}{\angstrom} absorption line
and the \ion{C}{4} \num{5801.35}, \SI{5811.97}{\angstrom} doublet, respectively.
Neither of these lines are expected to be seen in emission
from the photoionized gas in the nebula due to the high ionization potentials required.
They are therefore well-suited for tracing the diffuse dust-scattered starlight in the nebula.
It can be seen that the distribution of the two lines is very different.
For the \ion{He}{2} line, one finds \(W_\lambda\observed (4686) \approx \SI{0.2}{\angstrom}\)
across most of the nebula, but with consistently larger values to the SE,
with a pronounced peak of \(W_\lambda\observed (4686) = \SI{0.4 \pm 0.1}{\angstrom}\)
around the star \th2A.\@
In addition, there is a secondary peak of
\(W_\lambda\observed (4686) = \SI{0.3 \pm 0.05}{\angstrom}\),
located close to \th1D, the easternmost of the four Trapezium stars. 
In contrast, the \ion{C}{4} lines show a broad maximum of 
\(W_\lambda\observed (5801 + 5812) = \SI{0.16 \pm 0.04}{\angstrom}\),
centered to the SW of the Trapezium.

The most straightforward of these to interpret is \ion{C}{4},
since the 5801,5812 doublet is present in the spectrum of \th1C,
with a summed equivalent width \(W_\lambda^\star = \SI{0.44 \pm 0.04}{\angstrom}\)
\citep{Stahl:1996a}\footnote{The 10\% uncertainty is due to variation with rotational phase, see next paragraph.},
but is absent in all the other OB stars in the nebula
(as measured from the MUSE datacube).
The equivalent width observed in the nebula is smaller than this,
because of dilution by the nebular continuum (recombination plus 2-photon)
and the starlight from the lower mass members of the Trapezium cluster.
The fraction of the total observed continuum intensity that is due to
dust-scattering of \th1C can be estimated as \(f\scat = W_\lambda\observed / W_\lambda^\star\),
for which we find a maximum value of \(f\scat(\th1C;\,\ion{C}{4}) = 0.4 \pm 0.1\)
in the region between the Trapezium and Orion~S.\@

Although \th1C (spectral type O7V\,f?p\,var;
\citealp{Simon-Diaz:2006b, Maiz-Apellaniz:2019a})
is the hottest and most luminous star in the nebula,
its \ion{He}{2} 4686 absorption line is anomalously weak 
and varies markedly on a rotational timescale of \SI{15.4}{days}
\citep{Conti:1972a, Stahl:1993a}
because it is partially filled in by emission from the magnetosphere
and complex magnetically-channeled stellar wind
\citep{Donati:2002a}.
From Fig.~5 of \citet{Stahl:1996a}, it can be seen that \(W_\lambda^\star (\th1C;\,4686)\)
varies systematically from \SI{-0.1}{\angstrom} to \SI{0.4}{\angstrom}
as a function of rotational phase.%
\footnote{
  Negative equivalent widths correspond to net emission, rather than absorption.
}
Three-dimensional simulations \citep{ud-Doula:2013a} show that the line profile
is dominated by emission (\(W_\lambda^\star < 0\))
for viewing directions along the magnetic pole,
but by photospheric absorption (\(W_\lambda^\star > 0\))
for viewing directions in the magnetic equator.
Comparison with observed phase-dependent line profiles suggests that
both the line-of-sight from Earth and the magnetic dipole axis are inclined
at \(\approx \ang{45}\) from the rotation axis.
This implies that the spectrum seen from Earth,
integrated over the rotational period,
is broadly representative of the average stellar spectrum
seen by the surrounding nebula,
so that a value of \(W_\lambda^\star (\th1C;\,4686) \approx \SI{0.15}{\angstrom}\)
is appropriate, albeit with a large uncertainty. 

The second most luminous star in the nebula is the spectroscopic binary \th2A
(spectral type O9.2V + B0.5:V(n); \citealp{Maiz-Apellaniz:2019a}),
located to the SE of the nebula, beyond the Bright Bar.
This shows a much simpler \ion{He}{2} line profile
with \(W_\lambda^\star (\th2A;\,4686) = \SI{0.53}{\angstrom}\) \citep{Simon-Diaz:2006b},
which explains why \(W_\lambda\observed (4686)\) is higher around \th2A,
than around the Trapezium.
Applying the same argument as above yields a maximum value of
\(f\scat(\th2A;\,\ion{He}{2}) = 0.75 \pm 0.2\),
implying that the blue continuum in this area of the nebula is totally dominated
by scattered light from \th2A, which forms a roughly circular reflection nebula.

The same absorption line is present in the spectra of other Trapezium stars,
being strongest in the B0.5V star \th1D with \(W_\lambda^\star (\th1D;\,4686) \approx \SI{0.25}{\angstrom}\) \citep{Simon-Diaz:2006b}.
It is therefore interesting and suggestive
that the local peak in \(W_\lambda\observed (4686)\)
is on \th1D's side of the Trapezium.
This would not be expected if all the Trapezium stars lay
at the same distance from the scattering grains
because \th1C is 4 times brighter than \th1D
at visual wavelengths, so the latter star contributes relatively little
to the total luminosity of the Trapezium.
However, it has been suggested \citep{Smith:2005a} that \th1D may lie
closer to the background molecular gas,
roughly \SI{0.1}{pc} behind \th1C.
In such a case, it would be possible for \th1D to dominate the illumination
of a small reflection nebula around itself,
and the present results would tend to confirm that. 


\begin{figure}
  \centering
  \includegraphics[width=\linewidth]{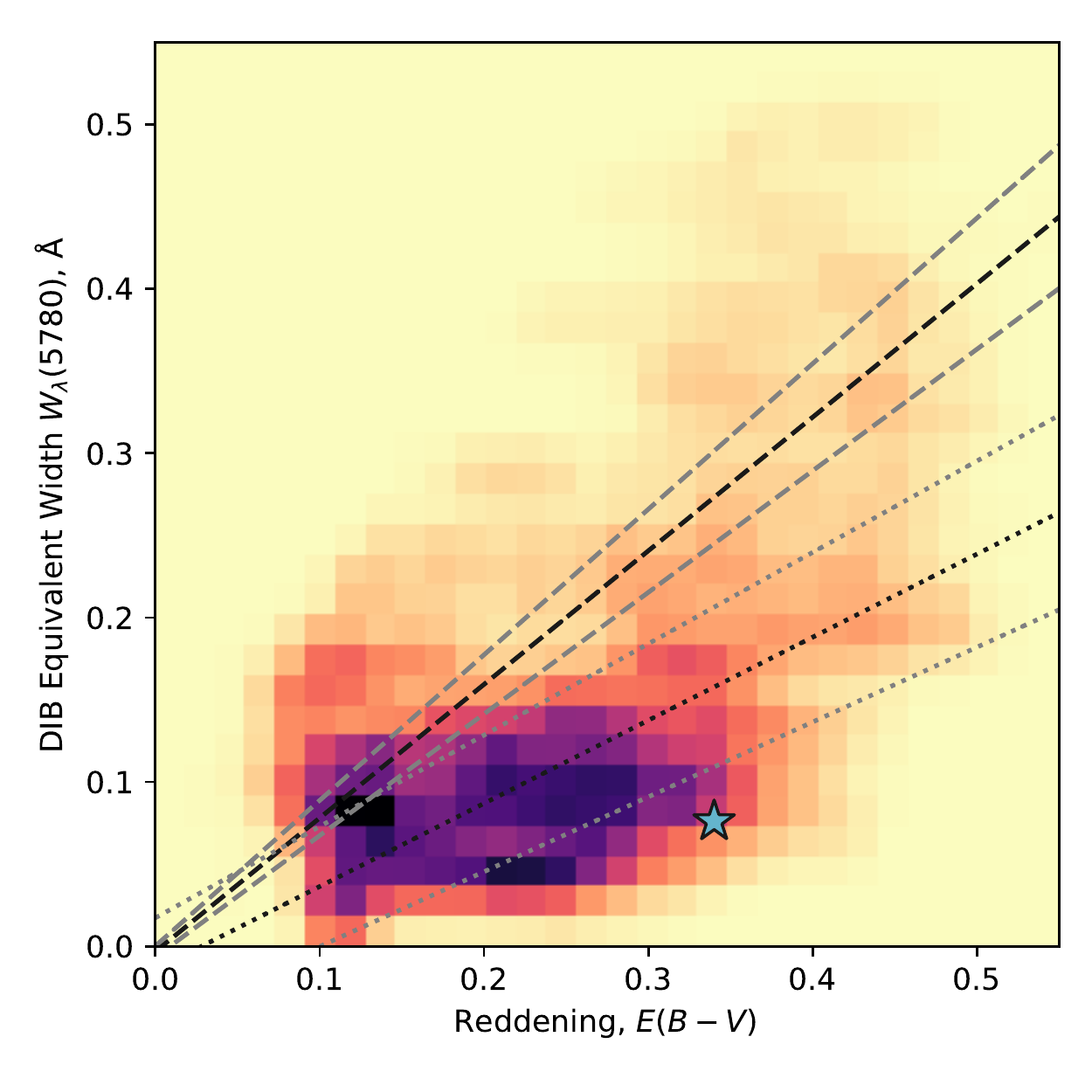}
  \caption{
    Correlation between DIB equivalent width,
    \(W_\lambda\observed(5780)\), and reddening, \(E(B - V)\).
    False-color image shows histogram of pixels across the face of the nebula,
    calculated from the MUSE spectral cube: 
    Fig.~\ref{fig:raman-multi-absorption-features} panel~c (\(y\) axis)
    and panel~f (\(x\) axis), assuming \(E(B - V) = 0.3845\, C(\hb)\).
    Straight lines shows fits to surveys of DIBs along \(\sigma\)-type sight lines
    to stars (dotted line, \citealp{Kos:2013a})
    and high-Galactic-latitude sight lines to external galaxies
    (dashed line, \citealp{Baron:2015a}).
    Star symbol shows result for \th1C \citep{Friedman:2011a}.
  }
  \label{fig:dib-redden}
\end{figure}

\subsubsection{Solid-state absorption features}
\label{sec:solid-state-absorpt}

A third type of absorption feature found in the nebular spectrum
is what is known as Diffuse Interstellar Bands (DIBs; \citealp{Heger:1922a}),
of which hundreds are now known \citep{Galazutdinov:2000a, Hobbs:2008a}.
The majority are believed to be due to large carbon-based molecules
associated with atomic gas \citep{Sonnentrucker:2014a}
but, apart from in a few instances \citep{Cordiner:2019a},
their exact nature is still uncertain \citep{Tielens:2014a, Omont:2019a, Lai:2020a}.
The strongest such feature found in the MUSE spectra is the well-known band
at \SI{5780}{\angstrom}, which is shown in Figure~\ref{fig:raman-multi-absorption-features}c.
For comparison, the foreground extinction to the nebula is shown in
Figure~\ref{fig:raman-multi-absorption-features}f,
which is calculated from the \ha{}/\hb{} ratio,
assuming the extinction curve of \citet{Blagrave:2007a}.

It is clear at a glance that the spatial distribution of the
DIB absorption and the continuum dust absorption are very similar.
This is not surprising, since many studies have established
a correlation between DIB equivalent widths and interstellar reddening
\citep{Friedman:2011a, Kos:2013a, Baron:2015a, Kreowski:2019b}.
This is tested quantitatively in Figure~\ref{fig:dib-redden},
which shows the correlation between \(W_\lambda\observed(5780)\)
and dust reddening \(E(B - V)\).
The relative weakness of the \SI{5797}{\angstrom} DIB feature in all
Orion Nebula sight lines is characteristic of \(\sigma\)-type DIBS
(after the prototype \(\sigma\)~Sco; \citealp{Krelowski:1988a}),
which tend to be found in atomic diffuse clouds. 
I therefore compare the results with a fit to 27 \(\sigma\)-type sight lines to
OB stars from the survey of \citet{Kos:2013a} (dotted line in Figure~\ref{fig:dib-redden}),
which covers the reddening range \(E(B - V) = 0.1\) to \num{1.2}.
I also show (dashed line in figure) a fit to over a million stacked external galaxy and quasar sight lines from SDSS \citep{Baron:2015a}, which cover the reddening range
\(E(B - V) = 0\) to \num{0.11}.
It can be seen that the nebular values follow the general trend of the fits,
albeit with considerable dispersion.
The tail of high \(W_\lambda\) above the fit lines may be due to the fact that
a significant fraction of the continuum is dust-scattered and those photons
will have traversed a larger column of dust than is measured by the foreground reddening.
In addition, the Balmer decrement will saturate at high column densities
if part of the dust is mixed with the emitting gas, resulting in an underestimate
of \(E(B - V)\).

A further DIB feature is seen in the spectra at \SI{6284}{\angstrom}.
However, it is hard to estimate the equivalent width of this feature
since it overlaps with telluric \chem{O_2} \(\gamma\)-band absorption
(see Figure~\ref{fig:raman-spectra-1d}).

\section{High-resolution spectroscopy of Raman-scattered \boldmath\ion{O}{1} \SI{1028}{\angstrom}}
\label{sec:keck-observations}

\begin{figure}
  \includegraphics[width=\linewidth]{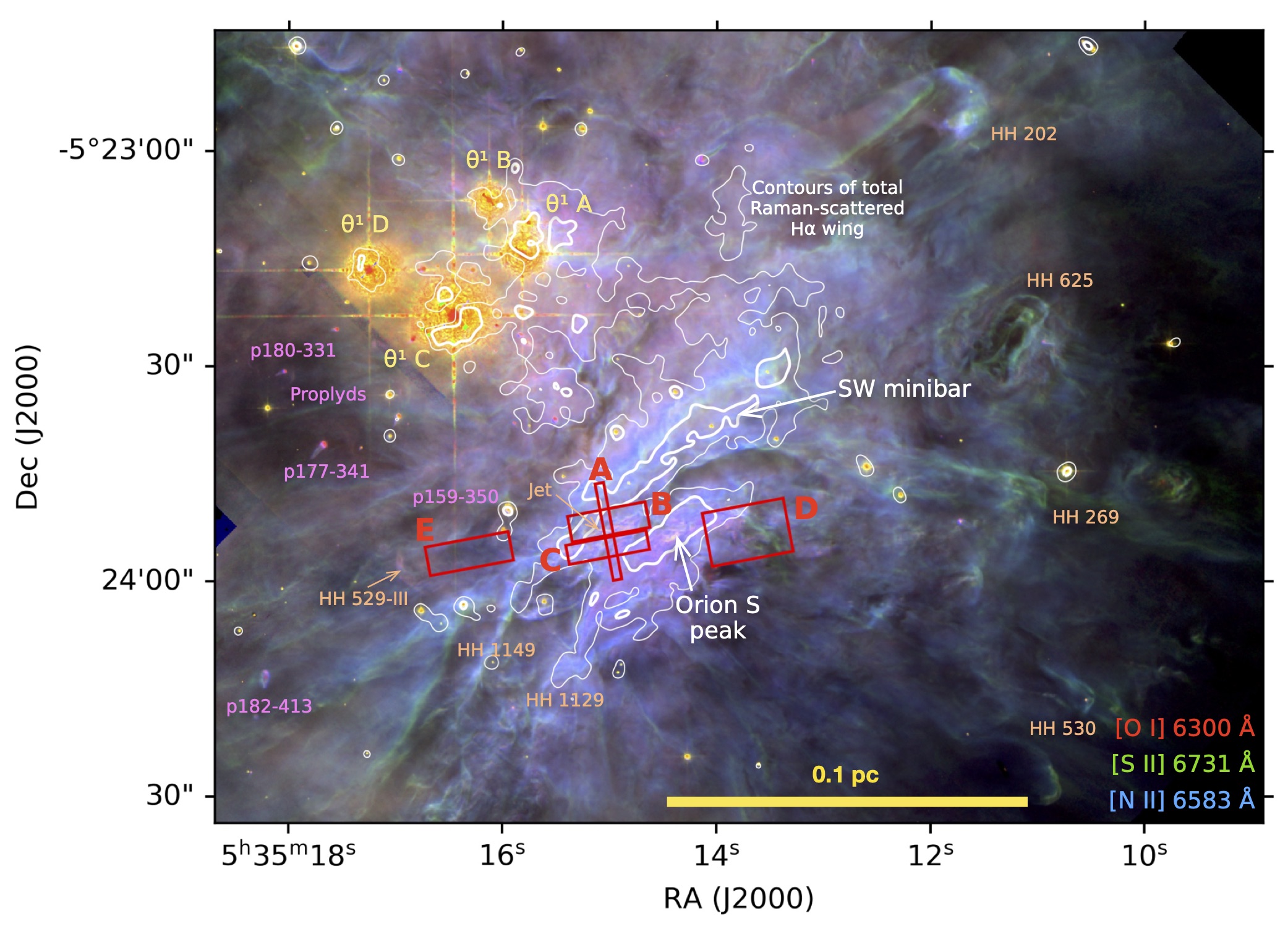}
  \caption{Zoom on Trapezium and Orion South region, showing the locations
    of extracted Keck spectra (red boxes).
    The background color composite image shows \textit{HST} WFPC2 observations
    in the filters F631N, F673N, and F658N \citep{Bally:2000a},
    which highlight emission from the ionization front and adjacent regions.
    White contours show the total Raman-scattered \ha{} wing emission
    (sum of the six wavelength bands listed in Table~\ref{tab:wav-bands}).
  }
  \label{fig:zoom-keck}
\end{figure}

\begin{figure*}
  \centering
  \includegraphics[width=\linewidth]{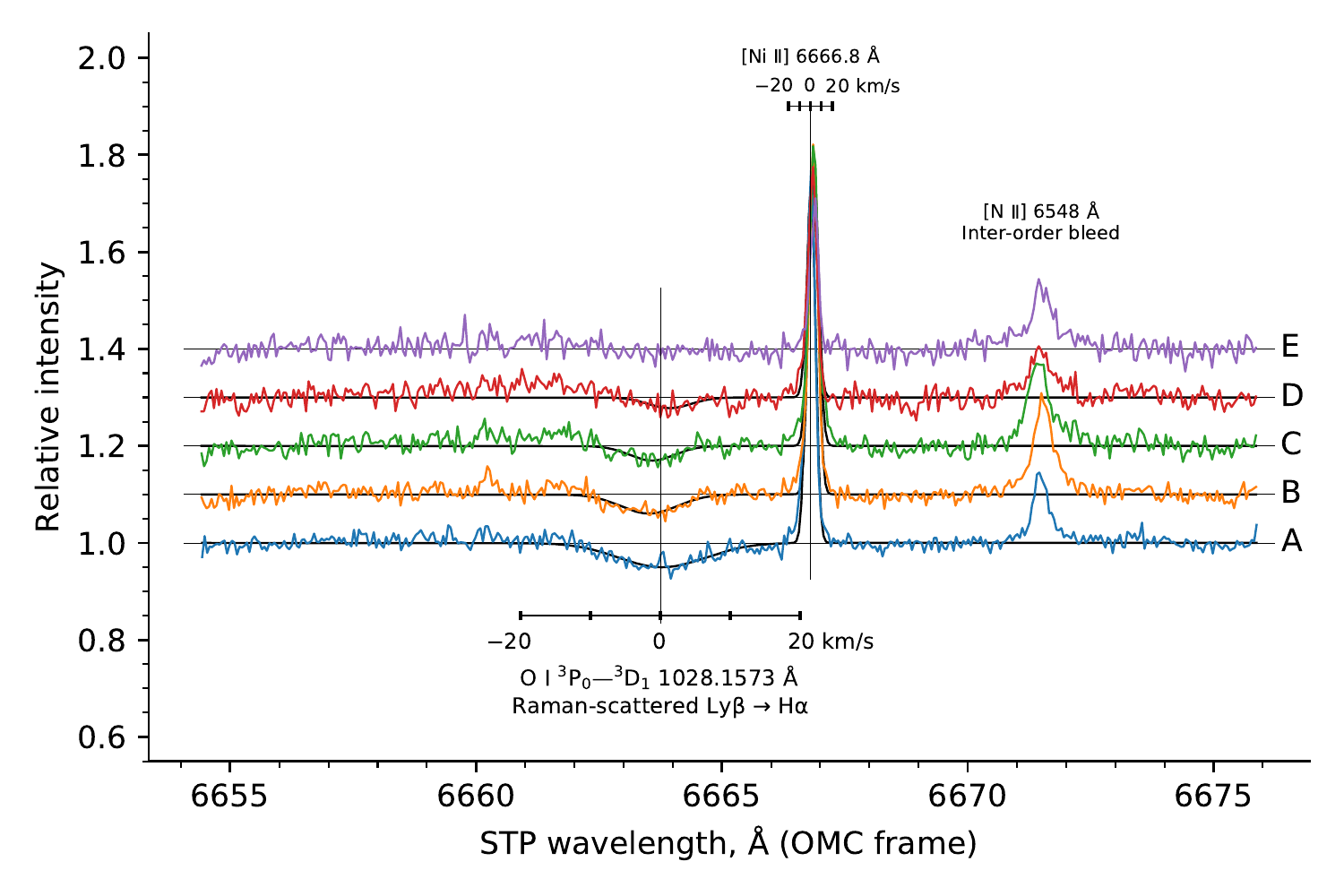}
  \caption{
    Continuum-normalized Keck HIRES spectra
    of Raman-scattered \ion{O}{1} absorption line for five regions in Orion~South.
    Spectra are offset vertically for ease of comparison.
    The location of each region is given in Figure~\ref{fig:zoom-keck}.
    Wavelengths are given on an air scale
    and in the rest-frame of the Orion Molecular Cloud, as
    defined by the peak velocity of \chem{^{13}CO}.
    Black continuous lines show single-Gaussian fits to the \ion{O}{1} absorption line
    and adjacent [\ion{Ni}{2}] emission line, with fit parameters given in Table~\ref{tab:line-fits}.
  }
  \label{fig:raman-keck}
\end{figure*}

In order to obtain observations of the Raman-scattered \ion{O}{1} absorption lines
at high spectral resolution, I have searched through archival spectra
of the Orion Nebula obtained with the HIRES spectrograph on the Keck~I telescope
\citep{Vogt:1994a}.
The observations, obtained in 1997 December~5 and 6,
were targeted at a small number of proplyds and jets in the nebula
and are described in detail in \citet{Henney:1999a} and \citet{Bally:2000a}.
Unfortunately, the stronger \SI{6633}{\angstrom} feature falls in a gap between adjacent spectrograph orders and so was not observed.
However, the weaker \SI{6664}{\angstrom} feature falls in order~53 and is detected in many of the spectra.
Results for the proplyds will be presented elsewhere,
and I concentrate here on 16 slit positions located in the Orion~South region,
roughly \SI{30}{arcsec} SW of the Trapezium.
Closely similar slit positions are co-added to improve S/N, 
forming five groups, A to E, with positions shown in Figure~\ref{fig:zoom-keck}.
Most of the slits are \SI{28}{arcsec} long and \SI{0.574}{arcsec} wide,
but only the central \SI{11}{arcsec} of order~53 is usable for faint lines
due to overlap of adjacent orders.\footnote{
  One slit in Group~A is \SI{14}{arcsec} long and does not suffer any order overlap.
}

The broad Raman-scattered wings to \ha{} are spread over several spectrograph orders
and uncertainties in the flat-field correction means that it is impossible to separate the wings from the nebular continuum.
Therefore, I simply fit a 5th-order polynomial to line-free sections of the spectrum
over the entire order (\SI{6649}{\angstrom} to \SI{6747}{\angstrom}) and divide
to give a continuum-normalized spectrum.
Results for the four groups are shown in Figure~\ref{fig:raman-keck} for the wavelength range of interest.
A shallow broad absorption feature is clearly seen around \SI{6664}{\angstrom} in regions A, B, and C,
and arguably in region~D, while no such feature is seen in region~E.\@
At this high spectral resolution (resolving power \(R \approx \num{50000}\)),
the absorption line is well-separated from the adjacent [\ion{Ni}{2}] emission line at \SI{6666.8}{\angstrom}.
From Figure~\ref{fig:zoom-keck} it can be seen that regions A, B, and C overlap with
contours of strong \ha{} wing emission in the SW minibar,
while region E lies in a higher ionization part of the nebula, where the \ha{} wing is weak.
This is consistent with the observed relative absorption depths,
assuming that the \SI{6664}{\angstrom} feature is due to Raman scattering of the \ion{O}{1} \SI{1028.15729} line
(see Table~\ref{tab:raman-wavelengths}). 

\begin{table}
  \caption{Parameters from single-Gaussian line fits to Keck HIRES spectra}
  \label{tab:line-fits}
  \begin{tabular*}{\columnwidth}{@{\extracolsep{\fill}} l  RRR}\toprule
    Region & \text{Relative amplitude},\ A & V,\ \si{km.s^{-1}} & \Delta V,\ \si{km.s^{-1}}  \\
    (1) & (2) & (3) & (4)  \\
    \midrule
    & \multicolumn{3}{c}{[\ion{Ni}{2}] emission \SI{6666.80}{\angstrom}} \\[2pt]
    A &  0.75 \pm 0.02 & 0.4 \pm 0.1 & 7.7 \pm 0.4\\
    B &  0.71 \pm 0.02 & 2.7 \pm 0.1 & 7.9 \pm 0.4\\
    C &  0.60 \pm 0.02 & 2.9 \pm 0.2 & 8.6 \pm 0.5\\
    D & 0.46 \pm 0.01 & 1.8 \pm 0.2 & 8.5 \pm 0.4\\
    E & 0.29 \pm 0.01 & 3.6 \pm 0.2 & 6.8 \pm 0.6\\
    \addlinespace[\medskipamount]
    & \multicolumn{3}{c}{\ion{O}{1} Raman-scattered absorption \SI{6663.75}{\angstrom}} \\[2pt]
    A & 0.050 \pm 0.003 & 0.3 \pm 0.4 & 14.5 \pm 1.0  \\
    B & 0.039 \pm 0.004 & -1.6 \pm 0.5 & 9.6 \pm 1.1  \\
    C & 0.030 \pm 0.005 & -1.2 \pm 0.6 & 7.6 \pm 1.5  \\
    D & 0.022 \pm 0.007 & 1.3 \pm 1.0 & 7.2 \pm 2.5    \\
    \bottomrule 
    \multicolumn{4}{@{}p{\columnwidth}@{}}{%
    \textsc{Columns}:
    (1)~Region of nebula covered by HIRES slits (see Figure~\ref{fig:zoom-keck}).
    (2)~Peak emission height or absorption depth relative to continuum.
    (3)~Centroid velocity in frame of molecular cloud. 
    (4)~Intrinsic full-width half-maximum line width,
    after correction for the instrumental width of \SI{6}{km.s^{-1}}.
    Note that in the case of the \ion{O}{1} Raman-scattered absorption line
    the centroid velocity and line widths have been divided by \num{6.4}
    to account for the stretching of the Doppler scale between the UV and optical frames.
    }
  \end{tabular*}
\end{table}

Table~\ref{tab:line-fits} shows results for each slit
from fitting both the \ion{O}{1} absorption line and the [\ion{Ni}{2}] emission line.
A single Gaussian component is used for each line
and the fits are performed with the \texttt{astropy.modeling} package.%
\footnote{\url{http://docs.astropy.org/en/stable/modeling/}}
The \ion{O}{1} ultraviolet rest wavelength and its optical counterpart after Raman scattering
are discussed in Appendix~\ref{sec:raman-theory},
while the [\ion{Ni}{2}] rest wavelength comes from hollow cathode tube observations \citet{Shenstone:1970a}.
Velocities are given in the reference frame of molecular gas in the Orion~South region,
as defined by the centroid of \chem{^{13}CO} emission \citep{Kong:2018a}:
\(V_{\mathrm{LSR}} = \SI{+9.0 \pm 0.2}{km.s^{-1}}\) or \(V_{\odot} = \SI{+28.1 \pm 0.2}{km.s^{-1}}\).
Both the emission and absorption lines have velocities that are close to that of the molecular gas,
as has been previously found for low-ionization lines in the nebula
\citetext{e.g., Fig.~14 of \citealp{Baldwin:2000a}}.\footnote{
  Although the [\ion{Ni}{2}] line, which likely arises near the ionization front,
  seems to show a consistent redshift of order \SI{2}{km.s^{-1}} with respect to CO,
  this is probably not significant given that the rest wavelength accuracy of \SI{0.03}{\angstrom}
  corresponds to \SI{1.4}{km.s^{-1}}.
}

The line width is of order \SI{8}{km.s^{-1}} for [\ion{Ni}{2}],
which is predominantly non-thermal
since the thermal broadening is only \SI{3}{km.s^{-1}} for \(T = \SI{e4}{K}\).
Compared with other optical emission lines that form near the ionization front,
the [\ion{Ni}{2}] width is intermediate between the widths of fluorescent lines
(e.g., [\ion{N}{1}]: \SI{6 \pm 3}{km.s^{-1}} \citealp{Ferland:2012a})
and collisionally excited lines
(e.g., [\ion{O}{1}]: \SI{11.5 \pm 2.6}{km.s^{-1}} \citealp{Garcia-Diaz:2008a}).
The Raman-scattered \ion{O}{1} line shows the greatest width
(\(W \approx \SI{15}{km.s^{-1}}\))
in region~A, which is where the absorption is strongest,
but is significantly narrower (\(W = \num{7}\) to \SI{10}{km.s^{-1}})
in the other regions.

In principle, the kinematics of Raman-scattered lines should be measurable
to a higher precision than that of regular lines.
This is because the stretching of the Doppler scale means that
the effective spectrograph resolving power is multiplied by \num{6.4} to become \(R' = \num{320000}\).
However, as can be seen from Table~\ref{tab:line-fits},
the precision of the centroid velocity and width measurements is actually \emph{lower}
for \ion{O}{1} than for [\ion{Ni}{2}].\footnote{%
  The uncertainties are estimated from the parameter covariance array
  returned by \texttt{astropy.modeling.fitting.LevMarLSQFitter},
  which uses a least-squares Levenberg-Marquardt algorithm implemented in
  \texttt{scipy.optimize.leastsq} \citep{Virtanen:2020a},
  which in turn is based on the \texttt{lmdif} and \texttt{lmder} routines
  in the MINPACK library \citep{More:1978a}. 
}
This is entirely due of the lower signal-to-noise of the absorption line measurements.
Therefore, much longer total exposure times
than the \(4 \times \SI{300}{s}\) of the slits that contribute to region~A
would be necessary to take full advantage of the boost in spectral resolution provided by Raman scattering.

\begin{figure}
  \centering
  \includegraphics[width=\linewidth]{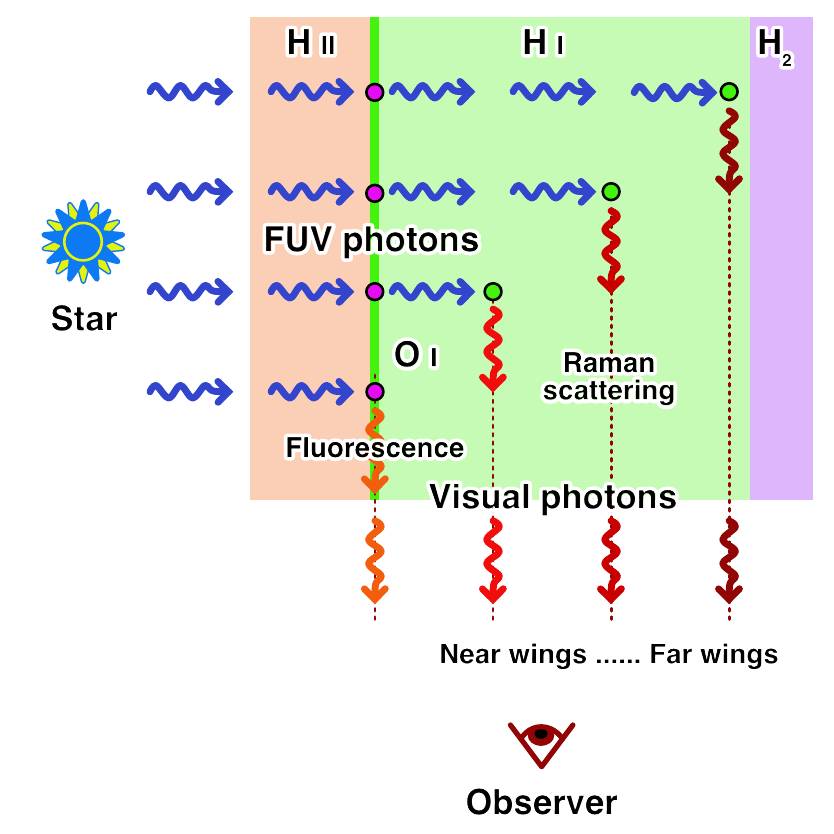}
  \caption{Simplified schematic of where Raman scattering occurs in a PDR.
    Stellar far-ultraviolet photons (blue wavy arrows)
    that are far-removed in wavelength from resonant \ion{H}{1} lines
    can pass through the photoionized \ion{H}{2} layer relatively unscathed.
    Some of them will be absorbed by metal resonance lines,
    such as \ion{O}{1} (magenta circles) close to the ionization front,
    and fluorescently converted to visual/IR photons (orange wavy arrows).
    This produces a spectral imprint of the \ion{O}{1} absorption
    on the FUV photons that propagate into the PDR.\@
    Once the FUV photons have traversed a sufficient column,
    they will be non-resonantly scattered by \ion{H}{1} (green circles)
    and a significant fraction will be Raman-converted to visual photons
    (red wavy arrows).
    For wavelengths farther from line center,
    this Raman scattering occurs deeper in the PDR
    and is eventually limited by competition from dust absorption
    and/or the finite total \ion{H}{1} column.
  }
  \label{fig:raman-scatter-cloud-schematic}
\end{figure}

\section{Discussion}
\label{sec:discussion}

In this section, I provide a provisional interpretation
of the observational evidence
from sections~\ref{sec:muse-spectr-mapp} and~\ref{sec:keck-observations}
in the light of simple heuristic models of Raman scattering,
such as is illustrated schematically in Figure~\ref{fig:raman-scatter-cloud-schematic}.
In section~\ref{sec:raman-radi-transf} I present some formal results
from radiative transfer theory that will be used in the following discussion.
Section~\ref{sec:broad-band-raman} investigates
how the spatial and spectral
characteristics of the broad \ha{} wings can be used to diagnose
the physical conditions in the PDR.\@
Section~\ref{sec:absorpt-feat-raman} considers 
what extra information
can be obtained from the narrow absorption features that are superimposed on the
Raman-scattered wings.

\subsection{Raman radiative transfer}
\label{sec:raman-radi-transf}

The Raman scattering process can be divided conceptually into three steps.
In step~1, radiative transfer in the FUV band determines the FUV mean intensity,
\(J_\lambda(\lambda_1)\), at some point in the nebula.
In step~2, the scattering itself (see Appendix~\ref{sec:raman-theory})
determines a visual-band emission coefficient \(j_\lambda(\lambda_2)\)
that is proportional to \(J_\lambda(\lambda_1)\), where the
visual wavelength, \(\lambda_2\),
and FUV wavelength, \(\lambda_1\),
are related by equation~\eqref{eq:wav-transform}.
In step~3, radiative transfer in the visual band determines the
observed intensity of the \ha{} wing, \(I_\lambda (\lambda_2)\).
I now consider each of these in turn. 

If the FUV radiation field at a point is dominated by a single star
of luminosity \(L_\lambda^*(\lambda_1)\) at a distance \(R\), then for step~1 we may write
\begin{equation}
  \label{eq:fuv-mean-intensity-from-star}
  4\pi J_\lambda(\lambda_1) = \frac{L_\lambda^*(\lambda_1)}{4 \pi R^2} \, e^{-\tau_1}
  \quad \si{erg.s^{-1}.cm^{-2}.\angstrom^{-1}}
\end{equation}
where \(\tau_1\) is the FUV optical depth between the point and the star:
\begin{equation}
  \label{eq:tau-fuv}
  \tau_1 = \int_0^R k_{\text{total}} (\lambda_1) \, dR'.
\end{equation}
The total FUV extinction coefficient can be written as a sum
over all absorbing species:
\begin{equation}
  \label{eq:k-fuv}
  k_{\text{total}} (\lambda_1) = \sum_i n_i \sigma_i(\lambda_1)
  \quad \si{cm^{-1}} 
\end{equation}
Potentially important broad-band absorbers are dust grains
and the \chem{H^0} Rayleigh/Raman scattering itself.
Resonance lines of ions, atoms, and molecules may also be important
over narrow ranges of \(\lambda_1\).
More realistically, one would sum over several equations like equation~\eqref{eq:fuv-mean-intensity-from-star} for the different stars,
and also include the diffuse field due to recombinations, Rayleigh scattering,
and dust scattering.

For step~2, assuming that the scattering is isotropic,%
\footnote{
  In reality, the scattering is not quite isotropic,
  but follows the Rayleigh scattering phase function:
  \(\frac34 \bigl( 1 + \cos^2\theta_{\text{scat}} \bigr)\),
  where \(\theta_{\text{scat}}\) is the angle between
  incident and scattered directions.
  However, this differs from the isotropic case by 50\% or less,
  so equation~\eqref{eq:raman-emissivity} is a reasonable approximation.
}
the visual-band emission coefficient is simply
\begin{multline}
  \label{eq:raman-emissivity}
  j_\lambda (\lambda_2) = \left(  \frac{\lambda_1}{\lambda_2}\right)^3
  J_\lambda(\lambda_1) \, n(\chem{H^0}) \, \sigma(\lambda_1) \, f_{\ha}(\lambda_1) 
  \\ \si{erg.s^{-1}.cm^{-3}.sr^{-1}.\angstrom^{-1}} ,
\end{multline}
where \(n(\chem{H^0})\) is the number density of neutral hydrogen atoms,
\(\sigma(\lambda_1)\) is the Rayleigh scattering cross section
in the \lyb{} wing given by
equation~\eqref{eq:total-cross-section-fit},
and \(f_{\ha}\) is the branching ratio for Raman conversion given by
equation~\eqref{eq:fha-fit}.
The factor \((\lambda_1 / \lambda_2)^3\) includes one power of \(\lambda_1/\lambda_2\)
to account for the reduction in energy per photon when passing from
the FUV to visual bands,
plus two powers of \(\lambda_1/\lambda_2\) to account for the transformation of
per-angstrom, \(d/d\lambda_1 \to d/d\lambda_2\).

For step~3, the observable intensity,
as shown in Figures~\ref{fig:raman-maps}ab and~\ref{fig:raman-bar-profile}a,
is given by an integral over all scattering points \(z\) along the line of sight (los):
\begin{equation}
  \label{eq:formal-solution}
  I_\lambda (\lambda_2) = \int_{\text{los}} j_\lambda(\lambda_2) \, e^{-\tau_2(z)} \, dz
  \quad \si{erg.s^{-1}.cm^{-2}.sr^{-1}.\angstrom^{-1}} ,
\end{equation}
where \(\tau_2\) is the visual-band optical depth between each point \(z\)
and the observer:
\begin{equation}
  \label{eq:tau-visual}
  \tau_2(z) = \int_0^z k_{\text{total}} (\lambda_2) \, dz'
\end{equation}
Similarly to the FUV case,
the visual-band extinction coefficient \( k_{\text{total}} (\lambda_2)\)
can be written as a sum over absorbing species,
but the only significant absorber in this case is dust.

\begin{figure*}
  \centering
  \includegraphics[width=\linewidth]{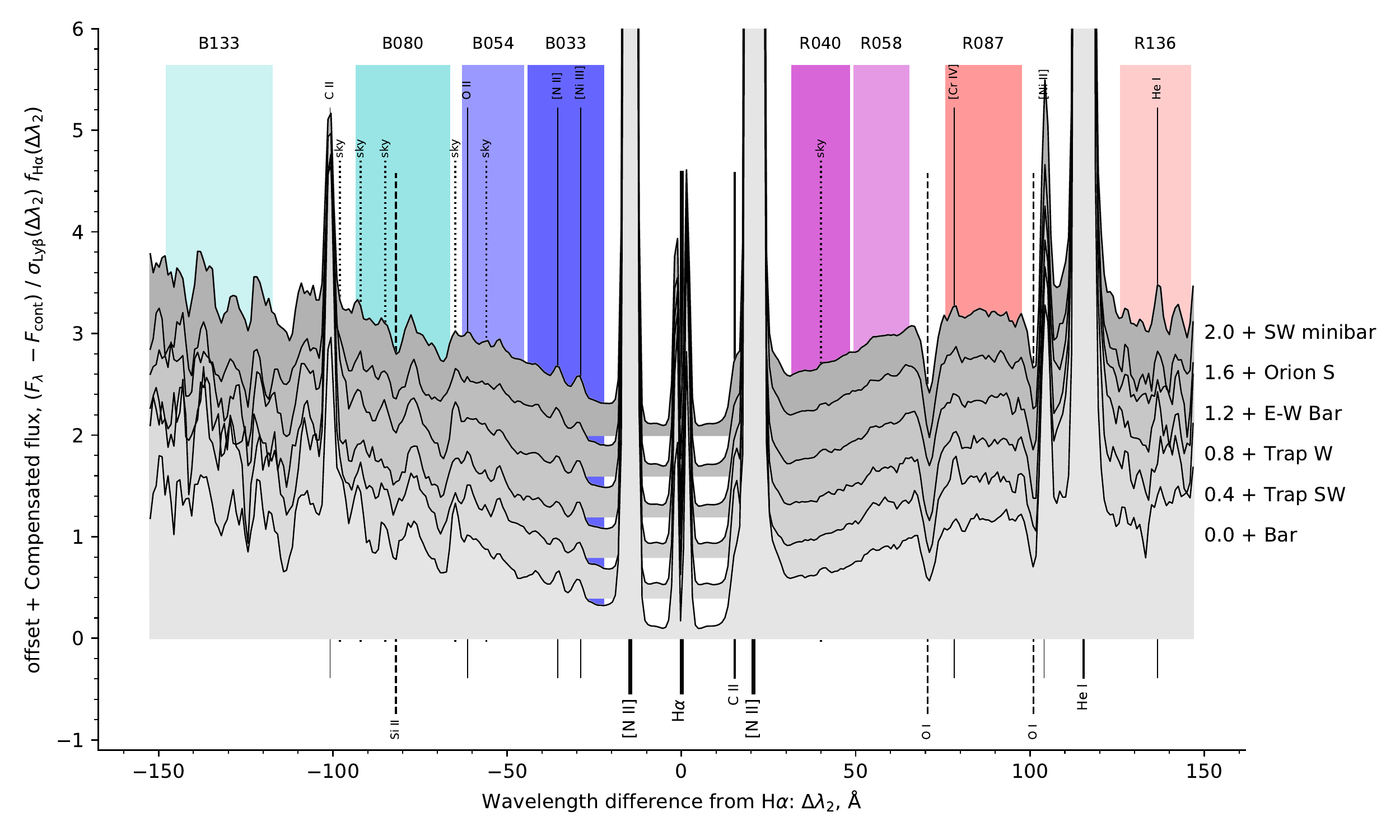}
  \caption{Same as Fig.~\ref{fig:raman-spectra-1d} but with
    \(F_\lambda\) divided by the effective cross section to Raman scattering
    \(\sigma\, f_{\ha}\) (see Appendix~\ref{sec:raman-theory})
    and zoomed in on the Raman wing
    wavelengths.  The spectrum is normalized by the average flux in
    the R087 band and each region is offset vertically by a constant
    value, as indicated at right.  In this presentation, optically
    thin Raman scattering of a flat FUV spectrum should give a
    constant value, which for the Orion data is only seen in the R087
    and R136 bands.}
  \label{fig:raman-compensated}
\end{figure*}

\subsection{Broad-band Raman scattering}
\label{sec:broad-band-raman}

An important question to answer with the Raman scattering is whether it is
optically thick or optically thin.
There are two optical paths to consider (see previous section):
the pre-scattered \(\tau_1\) of the FUV photons
and the post-scattered \(\tau_2\) of the visual-band photons.
Of these, it is \(\tau_1\) that is most interesting since it is affected
not only by dust extinction, but also by the opacity of the Rayleigh/Raman scattering itself.
The cross section of the \lyb{} wings varies by more than an order of magnitude
between the different Raman bands (column~6 of Table~\ref{tab:wav-bands}),
so it is possible that the nearer wings may be optically thick,
while the farther wings are optically thin. 

\subsubsection{Optically thin limit}
\label{sec:optically-thin-limit}

For the totally optically thin case, \(\tau_1, \tau_2 \ll 1\),
equations~(\ref{eq:fuv-mean-intensity-from-star},
\ref{eq:raman-emissivity},
and \ref{eq:formal-solution})
can be combined to yield
\begin{multline}
  \label{eq:optically-thin-intensity}
  I_\lambda (\lambda_2) \approx 
  \left(  \frac{\lambda_1}{\lambda_2}\right)^3 \, \sigma(\lambda_1) \, f_{\ha}(\lambda_1)
  \, \frac{L_\lambda^*(\lambda_1) } {16 \pi^2}
  \int_{\text{los}} \frac{n(\chem{H^0}) }{ R^2}  \, dz
  \\ \si{erg.s^{-1}.cm^{-2}.sr^{-1}.\angstrom^{-1}} .
\end{multline}
When considering the ratio of Raman-scattered intensities between two different wavelengths
for the same position on the nebula, then the line-of-sight integral cancels and one finds
\begin{equation}
  \label{eq:intensity-ratio-optically-thin}
  \frac{I_\lambda (\lambda_2)} {\sigma(\lambda_1) \, f_{\ha}(\lambda_1)} \propto L_\lambda^*(\lambda_1) .
\end{equation}
The left-hand side of this equation (approximately \((\Delta \lambda_2)^2\, I_\lambda (\lambda_2)\))
is plotted in Figure~\ref{fig:raman-compensated}
for the same extraction regions as in Figure~\ref{fig:raman-spectra-1d}. 
For parts of the spectrum where the scattering is optically thin,
this should be proportional to the incident FUV spectrum,
\(L_\lambda^*(\lambda_1)\),
which is dominated by the Trapezium OB stars. 

It can be seen that the red wing of this compensated spectrum
shows an approximately flat section
between \num{+80} and \SI{+150}{\angstrom} from \ha{},
covering the R087 and R136 Raman bands.
This is consistent with optically thin scattering of a flat FUV spectrum at these wavelengths.
However, the behavior is not so clear-cut in the blue wing,
where the compensated spectrum continues rising away from line center.
Although the slope is slightly shallower in the farther wing (B080 and B133 bands),
it never becomes as clearly flat as on the red side. 
The blue wing is noisier and more contaminated
by weak telluric and nebular lines, but this does not seem to be sufficient
to explain the discrepancy,
which can also be seen in Figure~\ref{fig:raman-spectra-1d} as a broad ``shoulder''
in the blue wing, extending from \num{6300} to \SI{6500}{\angstrom}.
One possibility is that the quadratic function used for continuum interpolation
is insufficiently accurate,
leading to a systematic overestimate of the strength of the far blue wing.
I have investigated the effect of changing the order of polynomial
used for the continuum interpolation,
and find that the intensity of the B080 and B133 bands are quite sensitive to this choice,
while the other bands are hardly affected.
This might be a result of intermediate scale structure
in the dust extinction law \citep{Whiteoak:1966a},
which has recently been shown to include a broad (\(\approx \SI{1000}{\angstrom}\))
feature centered at \SI{6300}{\angstrom} \citep{Massa:2020a}.

Closer to the line center, the results from the two wings are more consistent,
with both showing a reduction in compensated intensity.
This could mean one of two things: either the Raman wings become optically thick
for \(\vert \Delta\lambda_2 \vert < \SI{70}{\angstrom}\),
\emph{or} the incident FUV spectrum is no longer flat for \(\vert \Delta\lambda_1 \vert < \SI{1.7}{\angstrom}\),
which might correspond to the photospheric \lyb{} absorption line profile
in the spectra of the Trapezium stars.

\begin{figure}
  \includegraphics[width=\linewidth]{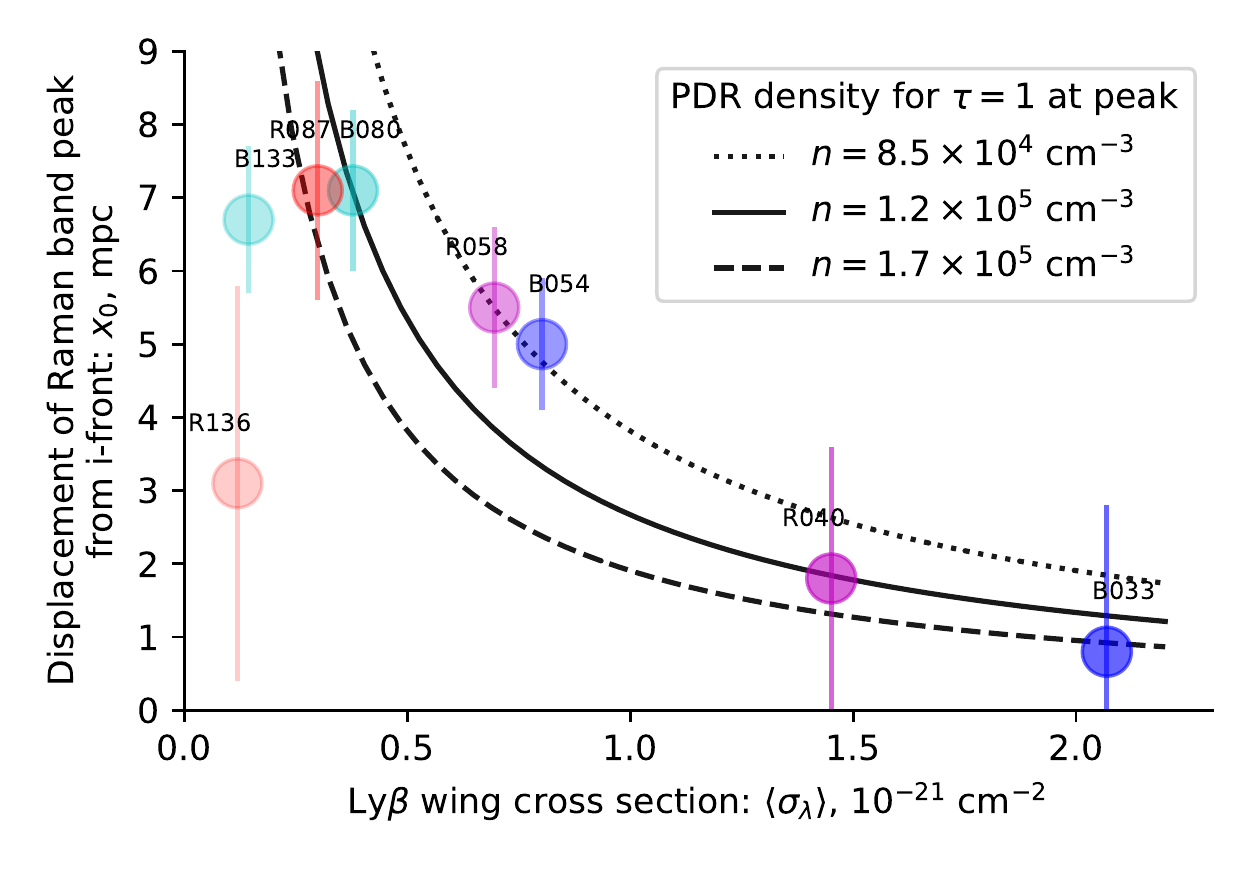}
  \caption{Displacement from the ionization front of the emission peak
    in each Raman band (column 3 of
    Table~\ref{tab:bar-profile-peaks}), plotted as a function of the
    cross-section of the \lyb{} wing (column 6 of
    Table~\ref{tab:wav-bands}).  Color scheme is as in
    Fig.~\ref{fig:raman-spectra-1d} with error bars showing 10\% of
    the peak FWHM (column 4 of Table~\ref{tab:bar-profile-peaks}).
    Black lines show a simple model in which the peaks occur at
    \(\tau = 1\) for 3 different constant neutral hydrogen densities, as
    indicated in the legend. }
  \label{fig:raman-band-displacements}
\end{figure}

\subsubsection{Optically thick limit}
\label{sec:optic-thick-limit}

An independent way of discriminating between optically thick and thin scattering
is to look at the spatial distribution of the Raman-scattered intensity.
In the optically thin limit, this will simply track the spatial distribution of scatterers
and so should not vary between the Raman bands.
In the optically thick limit, on the other hand, the scattered light will typically
arise at \(\tau_1 \approx 1\), as I will now demonstrate.

Consider a simplified plane-parallel geometry, as depicted in Figure~\ref{fig:raman-scatter-cloud-schematic}.
The FUV photons (blue wavy lines) propagate horizontally,
with \(\tau_1\) increasing towards the right.
The visual photons propagate vertically (red wavy lines), but I assume here that
either \(\tau_2\) is small or it is independent of \(\lambda_2\),
in which case it will have no \emph{differential} effect between the Raman bands.
If the \emph{total} FUV optical depth is very large,
then from equation~(\ref{eq:fuv-mean-intensity-from-star})
the mean value of \(\tau_1\) weighted by the FUV mean intensity is
\begin{equation}
  \label{eq:tau1-equals-unity}
  \vert \tau_1 \vert = \frac {\int_0^\infty \tau_1 \, e^{-\tau_1} \, d\tau_1}  {\int_0^\infty e^{-\tau_1} \, d\tau_1} = 1 , 
\end{equation}
where it is also assumed that \(R\) does not vary significantly with \(\tau_1\).
If, in addition, \(n(\chem{H^0})\) and the line-of-sight path length are roughly constant,
then from equations~(\ref{eq:raman-emissivity}, \ref{eq:formal-solution})
the mean \(\tau_1\) weighted by the observed intensity \(I_\lambda(\lambda_2)\) is also \(\vert \tau_1 \vert \approx 1\).

\subsubsection{Raman scattering as a diagnostic of
  the neutral hydrogen density in the PDR}
\label{sec:raman-scattering-as}

Based on the results of the previous section,
one would expect that the peak Raman emission should occur for a
\lyb{} wing optical depth of \(\tau_1 \approx 1\) for each band if the scattering is optically thick.
Disregarding dust opacity for the time being,
this corresponds to a displacement from the ionization front that varies as
\(x_0 \approx (n(\chem{H^0}) \langle \sigma_\lambda \rangle)^{-1}\), where \(n(\chem{H^0})\) is the neutral hydrogen
density in the PDR and \(\langle \sigma_\lambda \rangle\) is the \lyb{} wing cross section
for the FUV wavelengths that pump each band, which is given in
column~6 of Table~\ref{tab:wav-bands}.
In Figure~\ref{fig:raman-band-displacements}, this relationship is
compared with the measurements of the Orion Bar given in section~\ref{sec:emiss-prof-across}.
under the assumption that \(n(\chem{H^0})\) is constant over the depth range
\SIrange{2}{8}{mpc}.
It can be seen that a PDR density of order
\(n(\chem{H^0}) \sim \SI{1e5}{cm^{-3}}\) provides a good fit to
the observations for bands within about \SI{\pm 80}{\angstrom} of line center.
A systematic uncertainty of 10\% of the width of each emission peak is
assigned to the \(x_0\) values, which has the largest relative impact
on the near-wing bands: B033 and R040.  As a result, it is the B054
and R058 bands that are the most reliable indicators, yielding
\(n(\chem{H^0}) = \SI{8.5e4}{cm^{-3}}\) (dotted line).
This density is similar to that derived from the ratio of intensities
of far-infrared [\ion{C}{2}] \SI{158}{\micro m} and [\ion{O}{1}] \SI{145}{\micro m}
emission lines \citep{Bernard-Salas:2012a},
which trace slightly deeper layers of the PDR
(see Figure~\ref{fig:raman-bar-profile} and Table~\ref{tab:bar-profile-peaks}).
Densities derived from molecular hydrogen emission lines show a great deal of variation:
\citet{Luhman:1998a} find densities of \SI{e5}{cm^{-3}} to \SI{e6}{cm^{-3}},
whereas \citet{Kaplan:2017a} favor densities lower than \SI{e4}{cm^{-3}}.
It has been proposed that the Orion Bar PDR is highly inhomogeneous and clumpy
\citep{Burton:1990a}, but there is little evidence of clumpiness in the distribution
of Raman-scattered emission (Figure~\ref{fig:raman-maps}),
nor in the mid-infrared PAH and dust emission \citep{Salgado:2012a}.
This suggests that the shallow regions of the PDR,
with column densities up to \SI{e21}{cm^{-2}},
are relatively smooth.

The displacements of bands in the farther
wings fall below the predicted curve, significantly so for B133 and R136.
This is an indication that the farther bands may be optically thin instead of optically thick,
with the transition occurring near the wavelength of the B080 and R087 bands.
It is interesting that this is the same conclusion that is suggested
from analysis of the red wing of the compensated spectrum shown in Figure~\ref{fig:raman-compensated}, see section~\ref{sec:optically-thin-limit} above.
If this is the case, then it implies that the total \chem{H^0} column density through
the Orion Bar PDR is \(\approx \SI{2e21}{cm^{-2}}\).
However, an alternative interpretation could be that in these farther bands
the \lyb{} wing cross-section falls below the dust extinction cross-section,
so that most FUV photons are absorbed by dust before they can be Raman scattered.
In this scenario, the implication would be that \(\sigma_{\text{dust}} \approx \SI{5e-22}{cm^2.H^{-1}}\).
It is difficult to discriminate between these two possibilities,
especially since the dust opacity probably plays a role in \emph{determining}
the size of the \chem{H^0} column via the position of
the molecular hydrogen dissociation front \citep{Draine:1996a}.
The FUV dust absorption cross section implied by the Raman scattering is
within the range predicted for dense clouds \citep{Cardelli:1989a},
although \citet{Salgado:2016a} find that a lower value of
\(\sigma_{\text{dust}} \approx \SI{1.6e-22}{cm^2.H^{-1}}\) better reproduces the
distribution of PAH emission in the Orion Bar.
More detailed radiative transfer modeling is required to resolve the issue.

\subsection{Absorption features in the Raman pseudo-continuum}
\label{sec:absorpt-feat-raman}

The absorption lines seen in the Raman-scattered wings
are at STP visual-band wavelengths of \(\lambda_2 = 6633\) and \SI{6664}{\angstrom},
which correspond to vacuum FUV wavelengths of \(\lambda_1 = 1027\) and \SI{1028}{\angstrom}.
In Appendix~\ref{sec:raman-theory} it is shown that these can be identified as
resonant transitions from the ground state of atomic oxygen:
\(\ion{O}{1}\ \Term{2s 2p^4}{3}{P} \to \Term{2s 2p^3 3d}{3}{D^o}\).
The implication is that the line absorption occurs \emph{before} the
photon is Raman-scattered, and therefore contributes to \(\tau_1\)
(see section~\ref{sec:raman-radi-transf}).

The physical location of this \ion{O}{1} absorption is very well constrained
observationally because these very same resonance lines
(part of the UV~4 multiplet; \citealp{Moore:1976a})
are responsible for pumping some of the strongest
visual/IR fluorescent emission lines seen in the nebula.
Once excited to \Term{3d}{3}{D^o}, the oxygen atom will decay back to the ground level
with branching ratio 72.5\% via re-emission of an FUV photon,
or to the \Term{3p}{3}{P} level with branching ratio 27.5\% via emission of a near-infrared
\SI{1.129}{\micro m} photon (see Fig.~9 of \citealp{Walmsley:2000a}).%
\footnote{In addition, there are some semi-forbidden intercombination transitions
  that pump the optical forbidden lines,
  but their branching ratios sum to only 0.003\%.
}
From \Term{3p}{3}{P}, the only option is decay to \Term{3s}{3}{S^o}
via emission of \SI{8446}{\angstrom} and then back to ground
via emission of \SI{1304}{\angstrom}.
The spatial distribution of the \SI{8446}{\angstrom} emission line
(see Figures~\ref{fig:raman-maps}c and~\ref{fig:raman-bar-profile}c)
should therefore trace the location of the FUV \ion{O}{1} absorption.%
\footnote{
  The \Term{3p}{3}{P} level can also be populated via cascades that originate
  from pumping of the \(n \mathrm{s}\), \(n \mathrm{p}\), or \(n \mathrm{d}\)
  levels with \(n \ge 4\).
  Summing the intensities (in photon units) of all the shorter wavelength \ion{O}{1}
  fluorescent lines that I can find from the observations of
  \citet{Baldwin:2000a} and \citet{Esteban:2004a} and comparing with \SI{8446}{\angstrom}
  implies that 60\% of the pumping of \Term{3p}{3}{P} must come via \Term{3d}{3}{D^o}
  or \Term{4s}{3}{S^o}.  Given that the \SI{1.129}{\micro m} is slightly weaker than
  \SI{1.316}{\micro m} \citep{Walmsley:2000a},
  this means that about 25\% of the \SI{8446}{\angstrom} emission is caused by
  absorption of the UV~4 multiplet.}

It was shown in the section~\ref{sec:muse-spectr-mapp},
that this fluorescent emission arises very close to the ionization front in all cases
and tends to be arranged into filamentary structures that criss-cross the nebula
(red channel in Figure~\ref{fig:raman-maps}c),
of which the Orion Bar is the most prominent example.
As discussed in section~\ref{sec:lines-from-ionized},
the pumping should be most efficient at
a hydrogen column density of \(\approx \SI{e19}{cm^{-2}}\),
which is much smaller than the columns of \(> \SI{5e20}{cm^{-2}}\)
where the Raman scattering occurs. 

Note that it has long been recognised \citep{Grandi:1975a, Bautista:1999a}
that it is the stellar continuum radiation that excites the fluorescent \ion{O}{1} lines
in the Orion Nebula.
To the contrary, \citet{Dopita:2016a} propose that the excitation is due to the
\lyb{} \SI{1025.72}{\angstrom} emission line via the coincidence in wavelength
with the \SI{1025.76}{\angstrom} \ion{O}{1} line.
However, although this ``photoexcitation by accidental resonance'' (PAR) process
\citep{Kastner:1995a}
is well attested in novae and Herbig Ae/Be stars \citep{Mathew:2018a},
it requires the \ha{} transition to be optically thick,
which is never the case for \hii{} regions.
Powerful empirical arguments against the PAR process operating in the nebula are
(1)~that \SI{1.316}{\micro m} is stronger than \SI{1.129}{\micro m} \citep{Walmsley:2000a}
and (2)~the presence of lines originating from \ion{O}{1} levels as high as \Config{11s},
such as \SI{4980}{\angstrom} \citep{Esteban:2004a}.

\subsubsection{Absorption equivalent widths in the FUV frame}
\label{sec:impl-equiv-widths}

The visual-band absorption equivalent width \(W_\lambda\observed(6633)\)
is tightly correlated with
the relative brightness \(R\wing\) of the Raman wings (section~\ref{sec:cand-raman-scatt}).
Taking the ratio of these, one obtains
\(W_\lambda^{\text{Raman}}(6633) = W_\lambda\observed(6633) / R\wing\),
which is an equivalent width normalized to only the Raman pseudo-continuum itself,
rather than the full nebular continuum.
This is roughly constant across the nebula at
\(W_\lambda^{\text{Raman}}(6633) = \SI{2.2 \pm 1.0}{\angstrom}\)
(see Figure~\ref{fig:6633-Rwing}).
The equivalent width of the FUV absorption line is then smaller than
\(W_\lambda^{\text{Raman}}(6633)\) by a factor of
\((\lambda_2/\lambda_1)^2 \approx 41\), yielding \(W_\lambda^{\text{FUV}}(1027) \approx \SI{0.05}{\angstrom}\).
If the line profile is Gaussian, then the relation between velocity FWHM,
\(\Delta V\),
and equivalent width is \(W_\lambda = 1.064 A_0 \lambda_0 \Delta V / c\),
where \(A_0\) is the relative absorption depth at line center.
Assuming complete absorption (\(A_0 = 1\)), then yields \(\Delta V = \SI{15 \pm 6}{km.s^{-1}}\).
Very similar velocity widths were found for the \SI{6664}{\angstrom} absorption feature
from the high-resolution Keck spectra (Table~\ref{tab:line-fits}).

\subsubsection{The failure of rival explanations for the 6633, 6664~\AA{} lines}
\label{sec:rival-expl-6633}

However strong the case for identifying the absorption features at
\num{6633} and \SI{6664}{\angstrom} with Raman-scattered FUV
\ion{O}{1} absorption lines, it is nonetheless important to thoroughly
investigate potential competing explanations, if only to rule them
out.

\paragraph*{Why they cannot be dust-scattered stellar absorption lines}
\label{sec:cannot-be-stellar}

Various photospheric absorption lines of luminous O~stars
are visible in the spectrum of the diffuse nebula
due to scattering of the starlight by dust grains
in the \hii{} region and surrounding PDR.\@
These are clearest in the case of high-ionization lines
that are absent from the emission spectrum of the nebular gas,
such as the \ion{He}{2} and \ion{C}{4} lines that were analysed above
in \S~\ref{sec:stell-absorpt-lines}. 
At first glance, it might seem that the equivalent width map
of the \SI{6633}{\angstrom} feature
(Figure~\ref{fig:raman-multi-absorption-features}a)
is similar to that of the \ion{C}{4} absorption lines
(Figure~\ref{fig:raman-multi-absorption-features}e).
However, there are important differences:
the ridge of peak \ion{C}{4} absorption lies at about \num{20} to \SI{30}{arcsec}
from the Trapezium stars, close to the ``Trap SW'' region shown in
Figure~\ref{fig:raman-fov-regions}b,
whereas the \SI{6633}{\angstrom} absorption ridges lie farther from the Trapezium
(\num{50} to \SI{60}{arcsec})
in the ``SW minibar'' and ``Orion S'' regions.
This implies that the spatial distribution of scatterers is different
between the two cases.
This is exactly what would be expected under the Raman-scattering hypothesis
for \SI{6633}{\angstrom}, since the scatterers are neutral hydrogen atoms,
located in the PDR outside the \hii{} region,
whereas the stellar continuum is scattered by dust grains,
some of which will be inside the \hii{} region.%
\footnote{One piece of evidence for the presence of dust
  inside the \hii{} region is that
  the gas-phase abundances of nickel and iron are seen to be enhanced
  in photoionized Herbig-Haro outflows within the nebula \citep{Mesa-Delgado:2009b},
  consistent with dust destruction in shocks.
  Indirect evidence also comes from the dust-scattered redshifted wing of
  nebular emission lines \citep{Henney:1998a}.
}

An even stronger argument against a visual-band stellar origin is that
no absorption features at \num{6633} or \SI{6664}{\angstrom} are seen in the
luminous OB stars of the nebula.
The closest match is a pair of \ion{Fe}{1} features at \num{6634} and \SI{6664}{\angstrom},
which are seen in absorption in the photospheres of several K and M-type
T~Tauri stars in the Orion Nebula cluster.
One of the brightest and most centrally located examples is
V2279~Ori (Parenago~1869, JW~499, TCC~52), a visual binary
(spectral type M0.5 + M2; \citealp{Daemgen:2012a})
which is the central star of
the prominent proplyd 159-350 (HST~3; \citealp{ODell:1993a}).
However, the luminosity of this source is only \(\approx \SI{10}{\lsun}\)
and its brightness around \SI{6650}{\angstrom} is 200 times smaller than that of \th1C.
Furthermore, photospheric lines are weak in T~Tauri stars due to accretion veiling
and the equivalent widths of these absorption features are only
\(W_\lambda^\star \approx \SI{0.1}{\angstrom}\).
Therefore, even several such stars would contribute at most
\SI{0.001}{\angstrom}
to the equivalent width in the diffuse nebular spectrum,
which is far less than the observed values
\(W_\lambda\observed = \SI{0.1}{\angstrom}\) to \SI{0.5}{\angstrom}
(Figure~\ref{fig:6633-Rwing}).

\paragraph*{Why they are unlikely to be visual-band
  neutral/molecular absorption lines from the PDR}
\label{sec:cannot-be-abs}

The fact that the \num{6633} and \SI{6664}{\angstrom} features are not seen
in the spectra of the luminous stars, such as \th1C,
means that they cannot arise in the foreground neutral gas of the Orion veil
\citep{Abel:2004a, Abel:2019a}.
However, if some neutral or molecular species were abundant in the dense PDR behind the nebula,
but \emph{not} in the Veil,
then it is theoretically possible that it might imprint an absorption signature
on the starlight that initially enters the PDR but is then reflected out again
by dust scattering.
On the other hand,
such a scenario seems very far-fetched and I have been unable to find
any candidate absorption features at relevant wavelengths.

\paragraph*{Why they cannot be solid-state absorption features}
\label{sec:cannot-be-solid}

The same argument as in the previous paragraph also rules out any possibility
that foreground DIB absorption might be responsible for these features,
with the additional argument that the \SI{6633}{\angstrom}
equivalent width does not correlate with foreground extinction,
as the known DIB at \SI{5780}{\angstrom} does.
However, that leaves open the possibility of the contrived back-scattering scenario discussed above. 
There is a DIB at \SI{6632.85}{\angstrom} \citep{Galazutdinov:2000a},
which is consistent in wavelength with the \SI{6633}{\angstrom} feature
but it is only seen in highly-reddened stars
and is always at least 10 times weaker than the \SI{5780}{\angstrom} feature.
From Figure~\ref{fig:dib-redden} it is conceivable that up to half of the
\SI{5780}{\angstrom} absorption might arise during back-scattering from the PDR rather than in foreground material for some sight lines. 
This would imply an equivalent width \(W_\lambda < \SI{0.02}{\angstrom}\)
for the \SI{6632.85}{\angstrom} DIB,
which is again much smaller than the observed \(W_\lambda\observed(6633) = 0.1\)
to \SI{0.5}{\angstrom} (Figure~\ref{fig:6633-Rwing}).



\subsubsection{Possible detection of a \SI{6481}{\angstrom} absorption feature}
\label{sec:poss-detect-si6481}

A further absorption line is tentatively detected at \SI{6481}{\angstrom}
in the B080 Raman band (see Figure~\ref{fig:raman-spectra-1d}),
which could correspond to one of the components
of the \ion{Si}{2} \(\Term{3s^2 3p}{2}{P^o} \to \Term{3s^2 5s}{2}{S}\)
doublet (the other component, at \SI{6362}{\angstrom} is hidden by a much stronger
[\ion{O}{1}] emission line at \SI{6363}{\angstrom}).
This doublet (vacuum FUV wavelengths \num{1020.7} and \SI{1023.7}{\angstrom}) pumps the
visual-band fluorescent emission lines \num{5957.6} and \SI{5978.9}{\angstrom}.
The distribution of the \SI{5978.9}{\angstrom} line is significantly different from
the \ion{O}{1} fluorescent lines, being more similar to singly-ionized forbidden lines
such as [\ion{S}{2}] and [\ion{N}{2}].%
\footnote{
  The \ion{Si}{2} \SI{5957.6}{\angstrom} line is blended with
  \ion{O}{1} \SI{5958.4}{\angstrom}, so cannot be used.
}
This is what would be expected since \chem{Si^+}, unlike \chem{O^0}, has a significant
column inside the ionized gas.

The \SI{6481}{\angstrom} feature is considerably weaker than the other absorption features
and is located in a more complicated region of the spectrum,
which makes it difficult to make quantitative measurements.
Note that \citet{Dopita:2016a} claim evidence of a broad \emph{emission} bump
in the blue Raman wing,
centered on the \ion{Si}{2} \SI{6481}{\angstrom} wavelength.
However, I find no clear evidence for such a bump in the Orion MUSE or Keck spectra.








\section{Summary}
\label{sec:conclusions}

By analysing spatially resolved spectroscopic observations  of the Orion Nebula,
I have demonstrated that the broad Raman-scattered wings
of the \ha{} \SI{6563}{\angstrom} line
(\(\delta \lambda = \SI{\pm 150}{\angstrom}\))
are a useful diagnostic of the interaction of far-ultraviolet radiation with
atomic hydrogen in the environs of high-mass stars.
The principal conclusions of this study are:
\begin{enumerate}[1.]
\item Hydrogen Raman scattering of starlight from \lyb{} to \ha{} wings
  occurs in the neutral photo-dissociation region (PDR),
  located between the hydrogen ionization front and dissociation front.  
\item The inner Raman wings (\(|\delta \lambda | < \SI{80}{\angstrom}\)) of \ha{}
  are optically thick and allow
  the density of neutral hydrogen atoms to be determined in edge-on PDRs.
  For the case of the Orion Bar, I find \(n(\chem{H^0}) \approx \SI{e5}{cm^{-3}}\)
  for the shallow part of the PDR (depths of up to \SI{10}{mpc}).
\item The outer Raman wings are either optically thin or are limited by
  competition with dust absorption.
  In the first case, the total column density of neutral hydrogen must be 
  \SI{2e21}{cm^{-2}} in the Orion Bar.
  In the second case, the FUV dust absorption cross section must be
  \SI{5e-22}{cm^{2}.H^{-1}}.
\item Far-ultraviolet resonance lines of neutral oxygen imprint their absorption
  onto the stellar continuum as it passes through the ionization front.
  The subsequent Raman scattering of this continuum yields absorption lines
  in the red \ha{} wing
  at transformed wavelengths of \SI{6633}{\angstrom} and \SI{6664}{\angstrom}.
  This is a unique signature of Raman scattering, which allows it
  to be easily distinguished from other processes that might produce broad \ha{} wings,
  such as electron scattering or high-velocity outflows.
\item
  The widths of the Raman-scattered absorption lines are of order \SI{2}{\angstrom},
  which would correspond to a velocity width of \SI{100}{km.s^{-1}} in the visual band,
  but only \SI{15}{km.s^{-1}} in the FUV band where the lines were formed.
  This ``magnification'' of Doppler velocity scales by the Raman scattering process
  allows spectrographs to operate at a \(6.4\) times higher effective spectral resolution,
  but observations with better signal-to-noise 
  than are currently available are required in order to take full advantage of this.
\end{enumerate}

\section*{Acknowledgements}
I am grateful for financial support provided by
\foreignlanguage{spanish}{
  Dirección General de Asuntos del Personal Académico,
  Universidad Nacional Autónoma de México},
through grant
\foreignlanguage{spanish}{
  Programa de Apoyo a Proyectos de Investigación
  e Inovación Tecnológica}
IN107019.
Scientific software and databases used in this work include
SAOImage~DS9\footnote{\url{https://sites.google.com/cfa.harvard.edu/saoimageds9}} \citep{Joye:2003a},
the Atomic Line List\footnote{\url{https://www.pa.uky.edu/~peter/newpage/}} \citep{Van-Hoof:2018a},
the Cloudy plasma physics code\footnote{\url{https://nublado.org}}
\citep{Ferland:2017a},
SIMBAD and Vizier from Strasbourg Astronomical Data Center (CDS)\footnote{\url{https://cds.u-strasbg.fr}},
and the following 3rd-party Python packages:
numpy, astropy, matplotlib, seaborn, scikit-image, reproject.

\section*{Data availability}
\label{sec:data-availability}

The primary data underlying this article are two spectroscopic datasets.
First, integral field spectral mosaics of the inner Orion Nebula,
obtained with the MUSE spectrograph on the VLT,
which are available from \url{http://muse-vlt.eu/science/m42/}.
Second, longslit spectra of the Orion~S region,
obtained with the HIRES spectrograph on the Keck~I telescope,
which are available from the Keck Observatory Archive at
\url{https://koa.ipac.caltech.edu/cgi-bin/KOA/nph-KOAlogin}.
Additional supporting data from other observatories and archives
may be accessed via the references given in the main text.
All data reduction and analysis programs used in this paper,
together with documentation and research notes may be found at
\url{https://github.com/will-henney/dib-scatter-hii}.

\bibliography{raman-refs}

\appendix

\section{Raman scattering theory}
\label{sec:raman-theory}

\begin{table*}
  \caption{FUV/optical wavelength equivalencies for Raman scattering}
  \label{tab:raman-wavelengths}
  ~\\[-\baselineskip]
  \begin{tabular}{L L L L C R C C C}\toprule
    \text{Ion} & \text{Transition} & J_i \to J_k & \lambda_1,\ \si{\angstrom} & \wn_1,\ \si{cm^{-1}} & \Delta\wn,\  \si{cm^{-1}}& \wn_2,\ \si{cm^{-1}} & \lambda_2,\ \si{\angstrom} & \lambda_{\text{air}},\ \si{\angstrom} \\
    \midrule
    & & & \multicolumn{2}{c}{\dotfill\(\quad \lyb,\ n = 1 \quad\)\dotfill} & & \multicolumn{3}{c}{\dotfill\(\quad \ha,\ n = 2 \quad\)\dotfill} \\
    \addlinespace[2pt]
    \ion{H}{1} & n\Term{s}{2}{S} \to \Term{3p}{2}{P} & 1/2 \to 1/2, 3/2 & 1025.72220 & 97492.283 & 0.000 & 15233.329 & 6564.553 & 6562.740\\
    \addlinespace
    \ion{O}{1} & \Term{2s^2 2p^4}{3}{P}  \to \Term{2s^2 2p^3 (^4S) 3d}{3}{D^o} & 0 \to 1 & 1028.15729 & 97261.383 & -230.900 & 15002.429 & 6665.587 & 6663.747\\
                 & & 1 \to 1 & 1027.43139 & 97330.100 & -162.183 & 15071.146 & 6635.196 & 6633.364\\
                 & & 1 \to 2 & 1027.43077 & 97330.159 & -162.124 & 15071.205 & 6635.170 & 6633.338\\
                 & & 2 \to 1 & 1025.76339 & 97488.369 & -3.914 & 15229.415 & 6566.240 & 6564.427\\
                 & & 2 \to 2 & 1025.76276 & 97488.429 & -3.854 & 15229.475 & 6566.215 & 6564.401\\
                 & & 2 \to 3 & 1025.76170 & 97488.530 & -3.753 & 15229.576 & 6566.171 & 6564.358\\
    \addlinespace
    \ion{Si}{2} & \Term{3s^2 3p}{2}{P^o} \to \Term{3s^2 5s}{2}{S} & 1/2 \to 1/2 & 1020.6989 & 97972.086 &  +479.803 & 15713.132 & 6364.104 & 6362.345 \\
                 & & 1/2 \to 3/2 & 1023.7001 &  97684.859 &  192.576 & 15425.905 & 6482.602 & 6480.811 \\
    \bottomrule
  \end{tabular}
\end{table*}

When a photon is Raman-scattered from the vicinity of \lyb{} (UV
domain) to the vicinity of \ha{} (optical domain) its wavelength is
transformed from \(\lambda_1\) to \(\lambda_2\).  Intervals in frequency
(\(\nu = c/\lambda\)) or wavenumber (\(\wn = 1 / \lambda\)) space are conserved
between the two domains. For example the wavenumber displacement from
the \ion{H}{1} line center can be written in two ways:
\begin{equation}
  \label{eq:delta-wavnum}
  \Delta\wn = \wn_1 - \wn(\lyb) = \wn_2 - \wn(\ha) \ ,
\end{equation}
from which it follows that
\begin{equation}
  \label{eq:wav-transform}
  \lambda_2 = \left( \frac1{\lambda(\ha)} +\frac1{\lambda_1} - \frac1{\lambda(\lyb)}\right)^{-1} \ .
\end{equation}
The wavelengths \(\lambda(\lyb)\) and \(\lambda(\ha)\), together with their
corresponding wavenumbers, are given in
Table~\ref{tab:raman-wavelengths} (all wavelengths are on the vacuum
scale unless otherwise noted).  For both lines, a weighted average
over the \Level{3p}{2}{P}{1/2} and \Level{3p}{2}{P}{3/2} upper levels
is used, assumed to be populated according to their statistical
weights, with individual component wavelengths obtained from
Tab.~XXVIII of \citet{Mohr:2008a}. Note that the electric dipole
selection rules mean that only \Config{3p \to 2s} transitions contribute
to \ha{} in the Raman scattering context.  The wavelength is therefore
slightly shorter than the value obtained for the \ha{} recombination
line, which includes additional contributions from \Config{3s \to 2p}
and \Config{3d \to 2p}. The shift is of order \SI{-0.05}{\angstrom} or
\SI{-2}{km.s^{-1}} with respect to the Case~B results reported in
Tab.~6a of \citet{Clegg:1999a}.

Also listed in Table~\ref{tab:raman-wavelengths} are the Raman
transformations \(\lambda_1 \to \lambda_2\) for the rest wavelengths of transitions
between the ground \Term{2s^2 2p^4}{3}{P} term of neutral
\chem{^{16}O} and the excited \Term{2s^2 2p^3 3d}{3}{D^o} term. The
\ion{O}{1} data is obtained from highly accurate laser metrology
\citep{Ivanov:2008a, Marinov:2017a}, with a precision of
\SI{0.08}{cm^{-1}} or better.  The fine structure splitting between
the \(J_k\) levels of the excited term (\(\sim \SI{0.1}{cm^{-1}}\)) is
much smaller than that between the \(J_i\) levels of the ground term
(\(\sim \SI{100}{cm^{-2}}\)), so that the 6 transitions fall into 3
well-separated groups.  The three transitions from the lowest energy
\(J_i=2\) level are very close to \lyb{}
(\(\Delta\wn \approx \SI{4}{cm^{-1}}\)), whereas the two transitions from
\(J_i=1\) (\(\Delta\wn \approx \SI{162}{cm^{-1}}\)) and the single transition
from \(J_i=0\) (\(\Delta\wn \approx \SI{231}{cm^{-1}}\)) lie increasingly to the
red.  The corresponding wavelengths in the optical domain,
\(\lambda_2\), are therefore on the red side of \ha{}.  The final column of
the table uses STP refractive indices \citep{Greisen:2006a} to convert
\(\lambda_2\) to air wavelengths, \(\lambda_{\text{air}}\), for ease of comparison
with ground-based optical spectroscopy.  The resultant wavelength is
\SI{6663.747}{\angstrom} for the line from \(J_i = 0\), with an
uncertainty of about \SI{0.004}{\angstrom}, which is much smaller than
typical observational precision (for instance, \SI{0.07}{\angstrom}
for a very high resolution spectrograph with resolving power of
\(R = \num{e5}\)). The two lines from \(J_i = 1\), with a separation
of \SI{0.028}{\angstrom}, will always be blended in observations,
giving a mean wavelength of \SI{6633.347}{\angstrom} (assuming the
upper levels are distributed according to statistical weight
\(2 J_k + 1\)).  Similarly, the three lines from \(J_i = 2\) have a
mean wavelength of \SI{6564.386}{\angstrom}, but this is so close to
\ha{} (corresponding to a Doppler shift of \SI{+75}{km.s^{-1}}) that
it would be very difficult to observe.

The final section of Table~\ref{tab:raman-wavelengths} gives data for
a resonance doublet of \chem{Si^+} whose components lie a few
\si{\angstrom} to the blue of \lyb{}.  The shorter of the two
components is Raman-transformed to
\(\lambda_{\text{air}} = \SI{6362.35}{\angstrom}\), which unfortunately
coincides with the collisionally excited [\ion{O}{1}] line at
\SI{6363.78}{\angstrom}.  The longer component is transformed to
\SI{6480.81}{\angstrom} in a region that is clear of any strong
nebular lines.  Wavelengths for these lines are based on energy levels
from \citet{Martin:1983a}, with an estimated precision of
\SI{0.1}{cm^{-1}}, which gives an uncertainty in the optical
wavelengths \(\lambda_2\) of \SI{0.04}{\angstrom}.

The total cross-section for the off-resonance \Config{1s \to 3p}
transition in \chem{H^0} is calculated in \S~2 of \citet{Chang:2015a}
from second order time-dependent perturbation theory. Results are
presented in the upper panel of Figure~1 of that paper in terms of a
Doppler velocity factor \(\Delta V_1\), which in the notation of the
current paper is
\begin{equation}
  \label{eq:chang-DeltaV1}
  \Delta V_1 = c \left( \frac{\lambda_1}{\lambda(\lyb)} - 1 \right) \ .
\end{equation}
The observed Raman-scattered wings of \ha{} that are analyzed in this paper
are typically within \(\Delta\lambda_2 \sim \pm \SI{100}{\angstrom}\) of the line
core.  It is therefore convenient to define a dimensionless wavelength
in the optical domain as
\begin{equation}
  \label{eq:x-optical-def}
  x = \frac{\lambda_2 - \lambda(\ha)}{\SI{100}{\angstrom}} \ ,
\end{equation}
which corresponds to \(x \approx \Delta V_1 / \SI{714}{km.s^{-1}}\) in the FUV
domain.  The total cross section in the range \(0.4 < |x| < 2.3\) can
then be fit as follows:
\begin{equation}
  \label{eq:total-cross-section-fit}
  \frac{\sigma}{\SI{e-21}{cm^2}} =  
  \begin{cases}
    0.2186 x^{-2} - 0.0344 x^{-1} - 0.0054 & \text{if \(x < 0\)} \\
    0.2367 x^{-2} - 0.0187 x^{-1} + 0.0004 & \text{if \(x > 0\)} 
  \end{cases}
\end{equation}
Note that separate fits are given for the blue (\(x < 0\)) and red
(\(x > 0\)) wings of \ha{} since the cross section, although
approximately Lorentzian, is not exactly symmetric, being stronger on
the blue side (by about 10\% for \(x = \pm 1\)).

The fraction of all \Config{1s \to 3p} excitations that result in Raman
scattering to an optical photon is given by the branching ratio,
\(f_{\ha}\), with the remaining fraction, \(1 - f_{\ha}\), resulting
in elastic Rayleigh scattering in which the photon remains in the FUV
domain.  The results for \(f_{\ha}\) are also shown in Figure~1 of
\citet{Chang:2015a} and can be fit as follows in the range
\(|x| < 5\):
\begin{equation}
  \label{eq:fha-fit}
  f_{\ha} = 0.2238 + 0.0363 x + 0.0024 x^2 \ .
\end{equation}
The relative accuracy of all these fits is better than 1\% within the
stated range (corresponding to
\(\sigma \approx \text{\SIrange{e-22}{e-21}{cm^2}}\)), which is perfectly
adequate for the purposes of this paper.  Note that the branching
ratio increases with \(x\), which means that the product
\(\sigma f_{\ha} \) is stronger on the red side of \ha{}.

\bsp	
\label{lastpage}

\end{document}